\UseRawInputEncoding
\documentclass[a4paper,fleqn]{cas-dc}
\usepackage[T1]{fontenc}
\usepackage[utf8]{inputenc}

\usepackage[numbers]{natbib}
\usepackage{booktabs}
\usepackage{siunitx}

\usepackage{placeins}
\usepackage{graphicx}%
\usepackage{multirow}%
\usepackage{amsmath,amssymb,amsfonts}%
\usepackage{amsthm}%
\usepackage{mathrsfs}%
\usepackage[title]{appendix}%
\usepackage[table]{xcolor}
\usepackage{textcomp}%
\usepackage{manyfoot}%
\usepackage{booktabs}%
\usepackage{algorithm}%
\usepackage{algorithmicx}%
\usepackage{algpseudocode}%
\usepackage{bm}
\usepackage{makecell}
\usepackage{tikz}

\usepackage{listings}%

\begin{document}
\newcommand{\beq}{\begin{equation}}
\newcommand{\eeq}{\end{equation}}
\newcommand{\turbo}{{{\textsf{FF\,TURBO$^{\sf{TM}}$\,PLUS}\,}}}
\newcommand{\leap}{{{\textsf{FF\,LEAP$^{\sf{TM}}$}\,}}}
\newcommand{\edge}{Edge\,}
\newcommand{\sky}{Sky\,}
\newcommand{\ray}{Ray\,}
\def\ek  #1{\textcolor{red}{{[#1]}}{}}
\def\jm  #1{\textcolor{blue}{{[#1]}}{}}

\def\tsc#1{\csdef{#1}{\textsc{\lowercase{#1}}\xspace}}
\tsc{WGM}
\tsc{QE}


\let\WriteBookmarks\relax
\def\floatpagepagefraction{1}
\def\textpagefraction{.001}

\shorttitle{Mechanical resilience of ultra-low-density racing-shoe foams}    

\shortauthors{Jeremy McCulloch and Ellen Kuhl}  

\title [mode = title]
{Mechanical resilience of ultra-low-density racing-shoe foams}



%

\author[1]{Jeremy McCulloch}[orcid=0009-0003-1960-2124, linkedin=jeremy-mcculloch]

\cormark[1]


\ead{jmcc@stanford.edu}


\credit{Conceptualization, Methodology, Formal analysis, Software, Visualization, Writing - original draft, Writing - review \& editing}

\affiliation[1]{organization={Department of Mechanical Engineering and
Wu Tsai Human Performance Alliance, Stanford University},
            addressline={318 Campus Drive}, 
            city={Stanford},
            postcode={94305}, 
            state={CA},
            country={USA}}

\author[1]{Ellen Kuhl} [orcid=0000-0002-6283-935X, linkedin=ellen-kuhl-5a3661207]


\ead{ekuhl@stanford.edu}


\credit{Conceptualization, Funding acquisition, Writing - review \& editing}

\cortext[1]{Corresponding author}



\begin{abstract}
High-performance racing shoes rely on ultra-low-density elastomeric foams that undergo large, repeated deformations during running. Yet little is known about how their mechanical properties vary throughout the shoe or change with repeated use. Here, we characterize the midsole foam in an elite-level racing shoe from the heel, midfoot, and toe of a new shoe and a shoe worn for 300 miles. Microscopy reveals a characteristic pore length scale of 128$\pm$18\,$\mu$m and supports an approximately isotropic continuum description. We then quantify the mechanical response under tension, compression, and shear. Remarkably, despite 300 miles of real-world use, the foam retains its mechanical response across all three loading modes and shoe regions. Energy return remains largely unchanged, with values of 85–93\% in tension and compression and 64–71\% in shear. At the same time, we observe strong regional variations, with tensile and compressive stiffnesses 38–51\% lower in the toe than in the heel. The foam also exhibits a pronounced tension–compression asymmetry in Poisson's ratio. Together, these findings reveal a spatially structured and mode-dependent mechanical response that remains largely preserved after 300 miles of use. This mechanical resilience may extend the functional lifetime of racing shoes, with implications for runners, replacement recommendations, and sustainability.\\[3.pt]
Source code, data, and examples are available at
https://github.com/LivingMatterLab/CANN.
\end{abstract}



\maketitle
\section{Introduction}
In the past decade, the development of racing shoes that combine carbon-fiber plates and ultra-light elastomeric foams has revolutionized distance running at all levels \cite{bruvere_mechanisms_2025}. At the highest level, two runners ran sub-two-hour times at the 2026 London Marathon, breaking the men's marathon world record \cite{grivas_why_2026}. Both runners were wearing the Adidas Adizero Adios Pro Evo 3, a high-performance racing shoe with a carbon-fiber plate embedded in an ultra-low-density elastomeric foam \cite{grivas_why_2026}. This incredible accomplishment came nearly a decade after Eliud Kipchoge ran a sub-two-hour marathon in an unofficial race as part of the INEOS 1:59 Challenge, while wearing one of the early models of the Nike Vaporfly, less than two years after carbon-fiber-plated shoes became available to the public \cite{rosenberg_what_2022}. Ruth Chepng'etich shattered the women's world record for mixed-gender races when she ran the first sub-2:10 marathon while wearing the Nike Alphafly 3
\cite{mabe_changepoint_2024}. Since the introduction of carbon-fiber-plated shoes a decade ago, the women's mixed marathon world record has dropped by over five minutes and the men's marathon world record has dropped by over 3 minutes. These improvements are consistent with evidence that the original Nike Vaporfly shoes improved running economy by over 4\% \cite{hoogkamer_comparison_2018}, and that the latest high-performance running shoes improve running economy by over 3\% compared to previous carbon-fiber-plated shoes \cite{kuzmeski_data_2026}.

In addition to revolutionizing distance running at the highest levels, carbon-fiber-plated shoes have become commonplace in amateur races across the world. One market estimate placed annual sales of carbon-fiber-plated shoes at \$1.45 billion in 2025, or about 6 million units \cite{deep_market_insights_carbon_2026}. At the same time, the qualifying times for high-level amateurs have dropped considerably: In the Boston Marathon, the qualifying time has dropped by more than 12 minutes in the last decade; for comparison, the last 12-minute drop in qualifying time took four decades \cite{boston_athletic_association_qualify_nodate}. As these high-performance shoes become more effective at improving running economy \cite{kuzmeski_data_2026}, they enable an increasing number of professional and amateur runners to achieve their goals. 
{\it{But how durable are modern carbon-fiber-plated racing shoes, and how do their mechanical properties evolve with repeated use?}}
Even with their widespread adoption, this fundamental question remains largely unanswered.\\[6.pt] 
{\sffamily{\bfseries{Benefits of carbon-plated racing shoes.}}}
The primary benefit of carbon-fiber-plated shoes is that they improve performance by decreasing the metabolic cost of running at a given speed \cite{whiting_metabolic_2022}. Carbon-fiber-plated shoes may also affect recovery and injury risk. In particular, runners wearing carbon-fiber-plated shoes had decreased muscle damage and less reported muscle soreness after running a marathon compared to runners wearing standard racing shoes \cite{kirby_influence_2019}. However, their high cost can limit access for amateur athletes, especially if runners need to replace them frequently. It is therefore important to determine how repeated use affects the mechanical properties that govern the response of their ultra-light midsole foams. 
\begin{figure*}[t]
\centering
    \includegraphics[width=0.8\linewidth]{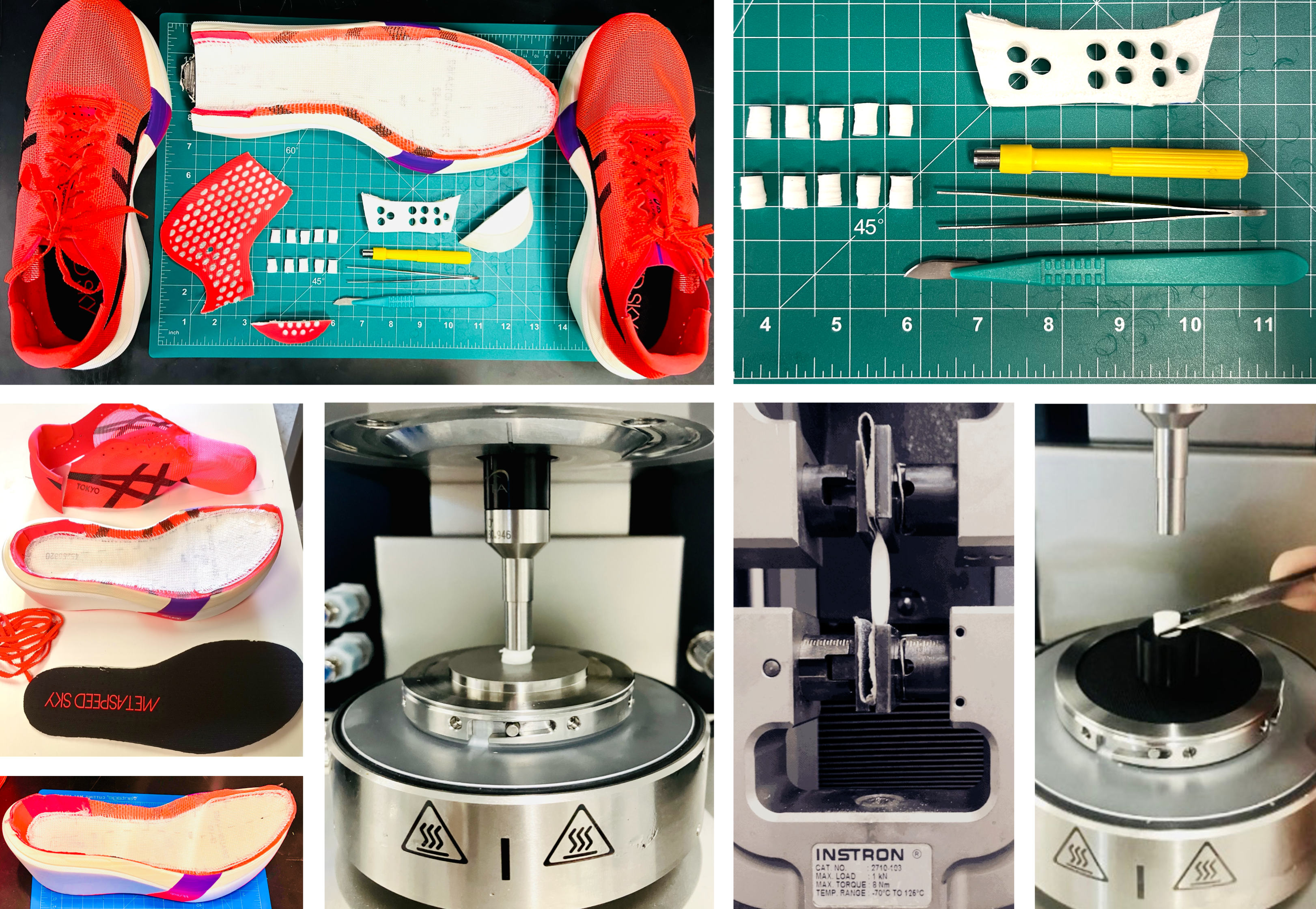}
\caption{{\sffamily{\bfseries{Sample preparation and mechanical testing of racing-shoe midsole foam.}}} Racing-shoe dissection and foam extraction from the toe, midfoot, and heel regions (left). Sample preparation for mechanical testing (top right). Compression, tension, and shear testing (bottom, from left to right). For each loading mode, we test $n=5$ samples from each region and quantify the stress–deformation response, transverse deformation, stiffness, and energy return.}
\label{fig00}
\end{figure*}
\\[6.pt]
{\sffamily{\bfseries{Durability of high-performance racing shoes.}}}
Running shoes are commonly recommended for replacement after approximately 300--500 miles, in part because repeated loading can reduce the resilience of their midsole foam. Yet scientific evidence for replacement thresholds in carbon-fiber-plated racing shoes remains limited, and most manufacturers provide no specific guidance on the mileage over which their performance benefits persist. Previous studies have characterized the durability of conventional running shoes, which typically use ethylene-vinyl acetate (\textsf{EVA}) in their soles \cite{wang_durability_2012, wang_changes_2010}.
Repeated loading of \textsf{EVA} midsole foams has been associated with changes in compressive response, density, energy absorption, and energy return, but these studies primarily rely on laboratory-based mechanical aging protocols to simulate accumulated running distance \cite{aimar_2023,lunchev_2022,verdejo_2004}. For carbon-fiber-plated shoes with polyether block amide (\textsf{PEBA}) foams, previous work reported a significant decrease in performance benefit after 450 km \cite{rodrigo-carranza_influence_2024}. However, new carbon-fiber-plated shoes use a wide variety of materials including polyether block amide (\textsf{PEBA}), thermoplastic polyurethane (\textsf{TPU}), and thermoplastic polyether ester elastomer (\textsf{TPEE}), with many shoes using multiple foams in a single shoe \cite{ou_multi-block_2026, rodrigo-carranza_influence_2024}. More broadly, footwear manufacturers tailor polymer composition and morphology 
to achieve application-specific mechanical properties \cite{fakayode_2025}. Replacing shoes too late may compromise their mechanical and performance characteristics, while premature replacement increases both cost and environmental impact \cite{cheah_manufacturing-focused_2013}. \\[6.pt]
{\sffamily{\bfseries{Properties of elastomeric foams.}}}
In our previous work, we showed that both \leap and \turbo \\
foams exhibit strongly nonlinear hyperelastic behavior with pronounced
tension--compression asymmetry and high energy return, two key
characteristics of modern racing-shoe foams
\cite{mcculloch_discovering_2026}. However, these measurements were limited
to new foams and did not quantify compressibility, spatial variations, or
the effects of repeated use. Here, we perform tension, compression, and shear
experiments while simultaneously measuring transverse deformation, and
compare three regions of new and worn elite racing shoes. Together, these measurements address three fundamental questions:
{\it{Do the mechanical properties, including energy return, persist after 300 miles of running? How strongly do these properties vary from heel to toe? And does the mechanical response depend on the loading mode?}}
\section{Methods}
\label{sec:methods}
In this study, we characterize \leap,
an ultra-low-density elastomeric foam
used in the ASICS Metaspeed racing shoe series. 
We test the foam in uniaxial tension, uniaxial compression, and simple shear \cite{mcculloch_discovering_2026}, and use a camera to measure the transverse strain during the tension and compression experiments. We test five samples from each of three regions, the heel, midfoot, and toe, in both a new shoe and a worn shoe. We compare a new unworn shoe with a shoe that an amateur marathon runner wore for 300 miles over seven months in training runs and races.
\subsection{Sample preparation}
We prepare the samples using 
a hot wire foam cutter 
(YaeTek Micromot Hot Wire ThermoCut Foam Cutting Machine, Yaemart, Duluth, GA) 
to cut along the bottom side of the carbon-fiber plate 
and separate it from the foam below \cite{mcculloch_discovering_2026}. 
We then cut this piece of foam 
into 10-mm-thick slabs along the length of the shoe. 
For uniaxial tension, 
we cut rectangular samples 
5\,mm thick, 10\,mm wide, and 50\,mm long.
For uniaxial compression and shear,
we create cylindrical samples 8\,mm in diameter and 10\,mm high 
using a biopsy punch.
We test samples from three regions:
toe, midfoot, and heel.
We define 
\textit{heel} samples as ${x}/{L} \in [0.1, 0.3]$, 
\textit{midfoot} samples as ${x}/{L} \in [0.4, 0.6]$, and 
\textit{toe} samples as ${x}/{L} \in [0.7, 0.9]$,
where $x$ is the distance from the heel
and $L$ is the total shoe length.
Before testing, 
we measure the exact dimensions of each sample using calipers.
\subsection{Microscopy} 
To characterize the microstructure of the shoe, we perform bright-field
microscopy of the new and worn samples from all three regions. We use a
surgical blade to cut thin slices of the samples and image them using a
laser-scanning Leica SP8 confocal microscope with a 10$\times$ 0.4-NA dry
objective. We capture a 12-bit image of each sample. We determine the
characteristic pore length scale from the radial spatial-frequency spectrum
of each image. To characterize microstructural anisotropy, we determine the
local in-plane pore-wall orientation $\theta$ from the image-intensity
gradients over the range $0^\circ\leq\theta<180^\circ$ and report the
normalized orientation distributions in $10^\circ$ intervals. We quantify
deviations from a uniform orientation distribution through the second-order
orientation parameter
$A_2=\left|\left\langle\exp(2{\rm i}\theta)\right\rangle\right|$,
where zero represents a uniform orientation distribution and one represents
perfect alignment.
\subsection{Mechanical testing}
\label{sec:mech-test-methods}
For each region, heel, midfoot, and toe, 
we test $n=5$ samples, 
both for the new shoe and the worn shoe,
in tension, compression, and shear 
\cite{mcculloch_discovering_2026}. For all tests, we choose a strain rate of 0.25/s, the largest possible strain rate that does not induce significant inertial effects. Rather than testing quasi-statically, we intentionally select a fast strain rate to reflect the loading rates during running, when ground-contact times are typically 100--200 ms \cite{chapman_ground_2012}. We ignore the first preconditioning cycle and estimate the stress during steady-state loading since the foam undergoes tens of thousands of loading cycles during a single race. \\[6.pt]
{{\sffamily{\bfseries{Tension.}}}}
For the tension tests,
we use an Instron 5848 
(Instron, Canton, MA) 
equipped with a 100\,N load cell. 
We clamp and preload the sample.
Then, we cyclically load the sample 
to a maximum stretch of $\lambda = 1.3$ 
at a rate of $\dot{\lambda} = 0.25$/s for six cycles.  \\[6.pt]
{{\sffamily{\bfseries{Compression.}}}}
For the compression tests, 
we use a Discovery HR20 Hybrid Rheometer 
(TA Instruments, New Castle, DE) 
equipped with a 10\,N load cell and an 8 mm tool. 
After preloading, 
we cyclically load the sample 
to a minimum stretch of $\lambda = 0.4$ 
at a rate of $\dot{\lambda} = 0.25$/s for six cycles.\\[6.pt]
{\sffamily{\bfseries{Shear.}}}
For the shear tests, 
we use a Discovery HR20 Hybrid Rheometer 
(TA Instruments, New Castle, DE) 
equipped with 
a custom 3D-printed end effector and a platform 
with sandpaper glued to both the end effector and the platform.
After preloading, 
we compress the sample
to a stretch of $\lambda = 0.8$ 
at a rate of $\dot{\lambda} = 0.25$/s. 
Then, we apply a sinusoidal angular displacement 
with a maximum shear strain of $\gamma = 0.25$ 
and a maximum shear strain rate of $\dot{\gamma} = 0.25$/s, 
which corresponds to an angular frequency of 1\,rad/s. 
\subsection{Video processing}
\label{sec:methods_video}
To quantify the transverse strain, we record a video of each experiment. We mount an iPhone 17 on a tripod, 25\,cm from the sample, and capture videos at a resolution of 4K, a frame rate of 60 frames/s, and 2$\times$ zoom, with a dark background to ensure maximal contrast for automated video processing.  
To compute the transverse stretch as a function of time, 
we construct an automated video analysis pipeline 
to compute the width and height of the sample in each frame of the 
tension and compression videos. 
First, we identify which pixels of each frame correspond to the sample. 
Then, we use this segmentation to quantify the width and height of the sample in pixels. 
Finally, we process the data from all frames to obtain a single curve that maps 
axial stretch to transverse stretch for each experiment. 
We implement these steps in OpenCV and outline them in detail in the following sections.\\[6.pt]
{\sffamily{\bfseries{Segmentation.}}}
To segment the images, we generate a mask of all white pixels by thresholding the pixels with high brightness and low saturation, which we call the \textit{original mask}. We then erode this mask by $n_{\rm erode} = 10$ pixels and dilate it by $n_{\rm dilate} = 20$ pixels to remove any spurious islands of pixels without affecting the size of the sample. We then identify the largest connected component and find its bounding box. This process cuts off the corners of rectangles; thus, to preserve the initial shape of the sample, we take the intersection of the bounding box of the largest connected component and the original mask, which we call the \textit{final mask}. \\[6.pt]
{\sffamily{\bfseries{Dimension estimation.}}}
To estimate the \textit{sample height} $h(t)$ from the final mask, we use the height of the bounding box. 
To estimate the \textit{sample width} $w(t)$, we use the width of the sample at its vertical midpoint. 
To do so, we first crop the final mask to the middle vertical third of its bounding box. Next, we fit a parallelogram to the resulting area. 
Finally, we compute the distance in pixels between the left and right sides of this parallelogram, which we use as our estimate of the sample width. \\[6.pt]
{\sffamily{\bfseries{Postprocessing.}}}
By applying segmentation and dimension estimation to each frame of each video, we obtain the width and height of the sample in pixels as a function of time. To convert this to a single function that maps axial stretch to transverse stretch, we first find the relative maxima and minima of the height $h(t)$ and use these to identify the start and end of each loading and unloading interval $t_0$, $t_1$, $t_2$, etc. Then, we compute the axial and transverse stretches $\lambda_1(t)$ and $\lambda_2(t)$ 
by dividing the current height and width 
$h(t)$ and $w(t)$
by the initial height and width at the beginning of the first loading interval
$h(t_0)$ and $w(t_0)$,
$$
\lambda_1(t) = \frac{h(t)}{h(t_0)}
\quad \mbox{and} \quad
\lambda_2(t) = \frac{w(t)}{w(t_0)}\,.
$$
Finally, for each loading and unloading curve, we compute a function $f$ such that $\lambda_2(t) = f(\lambda_1(t))$ and average these functions across all loading and unloading curves. 
\subsection{Data processing}
\label{sec:methods_data_proc}
We first convert the raw load-displacement recordings 
into stress-stretch measurements.
In tension and compression, 
this results in data pairs $\{\lambda_1,P_{\rm{11}}\}$,
where 
the stretch $\lambda_1 = {h(t)}/{h(t_0)}$ 
is the ratio 
of the deformed height $h(t)$ to the initial height $h(t_0)$,
and the Piola stress 
$P_{\rm{11}} = {F}/{A(t_0)}$
is the axial force $F$ divided by the undeformed cross-sectional area $A(t_0)$.
In shear, 
this results in data pairs $\{\gamma,P_{\rm{12}}\}$,
where 
the shear strain $\gamma = \phi \, {r(t_0)}/{h(t_0)}$ 
is the torsion angle $\phi$ 
scaled by the ratio 
of the initial radius $r(t_0)$ to the initial height $h(t_0)$,
and we approximate the Piola stress
$P_{\rm{12}} = 2T / (\pi r^3(t_0))$
as the torque $T$
scaled by the inverse of the initial radius cubed $r^3(t_0)$.
For each testing mode---tension, compression, and shear---we average the loading and unloading curves for each sample to approximate the elastic stress as a function of stretch. 

For each sample and each loading mode, we also compute the energy output per cycle as the area under the unloading curve and the energy input per cycle as the area under the loading curve, and take their ratio to obtain the  \textit{energy return} for each sample
\cite{mcculloch_discovering_2026}.
In addition, we estimate the linear elastic moduli, shear moduli, and Poisson's ratios for each sample using linear regression \cite{mcculloch_discovering_2026}, 
$$E_{\rm ten} 
= \frac
 {\sigma_{\rm ten}\cdot\varepsilon_{\rm ten}^{\rm axial}}
 {\varepsilon_{\rm ten}^{\rm axial} \cdot \varepsilon_{\rm ten}^{\rm axial}}
  \quad \mbox{and} \quad
  E_{\rm com} 
= \frac
 {\sigma_{\rm com}\cdot\varepsilon_{\rm com}^{\rm axial}}
 {\varepsilon_{\rm com}^{\rm axial} \cdot \varepsilon_{\rm com}^{\rm axial}}
$$
$$
  G_{\rm shr} 
= \frac
 {\tau_{\rm shr}\cdot\gamma_{\rm shr}}
 {\gamma_{\rm shr} \cdot \gamma_{\rm shr}}$$
$$\nu_{\rm ten} 
=-\frac
 {\varepsilon_{\rm ten}^{\rm trans}\cdot\varepsilon_{\rm ten}^{\rm axial}}
 {\varepsilon_{\rm ten}^{\rm axial} \cdot \varepsilon_{\rm ten}^{\rm axial}}
  \quad \mbox{and} \quad
 \nu_{\rm com}
=-\frac
 {\varepsilon_{\rm com}^{\rm trans}\cdot\varepsilon_{\rm com}^{\rm axial}}
 {\varepsilon_{\rm com}^{\rm axial} \cdot \varepsilon_{\rm com}^{\rm axial}},$$
where the dot denotes the inner product over all data points within the fitting range,
$\varepsilon_{\rm ten}^{\rm axial} = (\lambda_1 -1)$ and 
$\varepsilon_{\rm com}^{\rm axial} = (\lambda_1 -1)$ 
are the axial engineering strains,
$\varepsilon_{\rm ten}^{\rm trans} = (\lambda_2 -1)$ and 
$\varepsilon_{\rm com}^{\rm trans} = (\lambda_2 -1)$ 
are the transverse engineering strains, and 
$\gamma_{\rm shr} = \gamma$ is the shear strain, and 
$\sigma_{\rm ten}$, $\sigma_{\rm com}$, and $\tau_{\rm shr}$ 
are the associated Cauchy stresses. We use axial and shear strains up to 10\% to fit the elastic moduli and axial and transverse strains up to 30\% to fit Poisson's ratio. We adopt a larger strain range for Poisson's ratio because the lateral strain data are quite noisy. 
Finally, we compute the \textit{strain energy at maximum deformation} for each loading mode by integrating the Piola stress over the strain and shear ranges, 
$$\psi_{\rm ten} = \mbox{$\int_{1.0}^{1.3}$} P_{11}(\lambda)d\lambda
  \quad \mbox{and} \quad
  \psi_{\rm com} = \mbox{$\int_{0.4}^{1.0}$}-P_{11}(\lambda)d\lambda$$ 
$$\psi_{\rm shr} = \mbox{$\int_{0.00}^{0.25}$} P_{12}(\gamma)d\gamma$$
Unlike the elastic modulus, this quantity depends on the material behavior across the entire loading path, rather than only on the small-strain regime. As a result, it is less sensitive to noise and outliers. In our analysis we use strain energy at maximum deformation to compare the elastic stress response of different shoes and regions. 
\begin{figure*}[t]
\centering
    \includegraphics[width=0.8\linewidth]{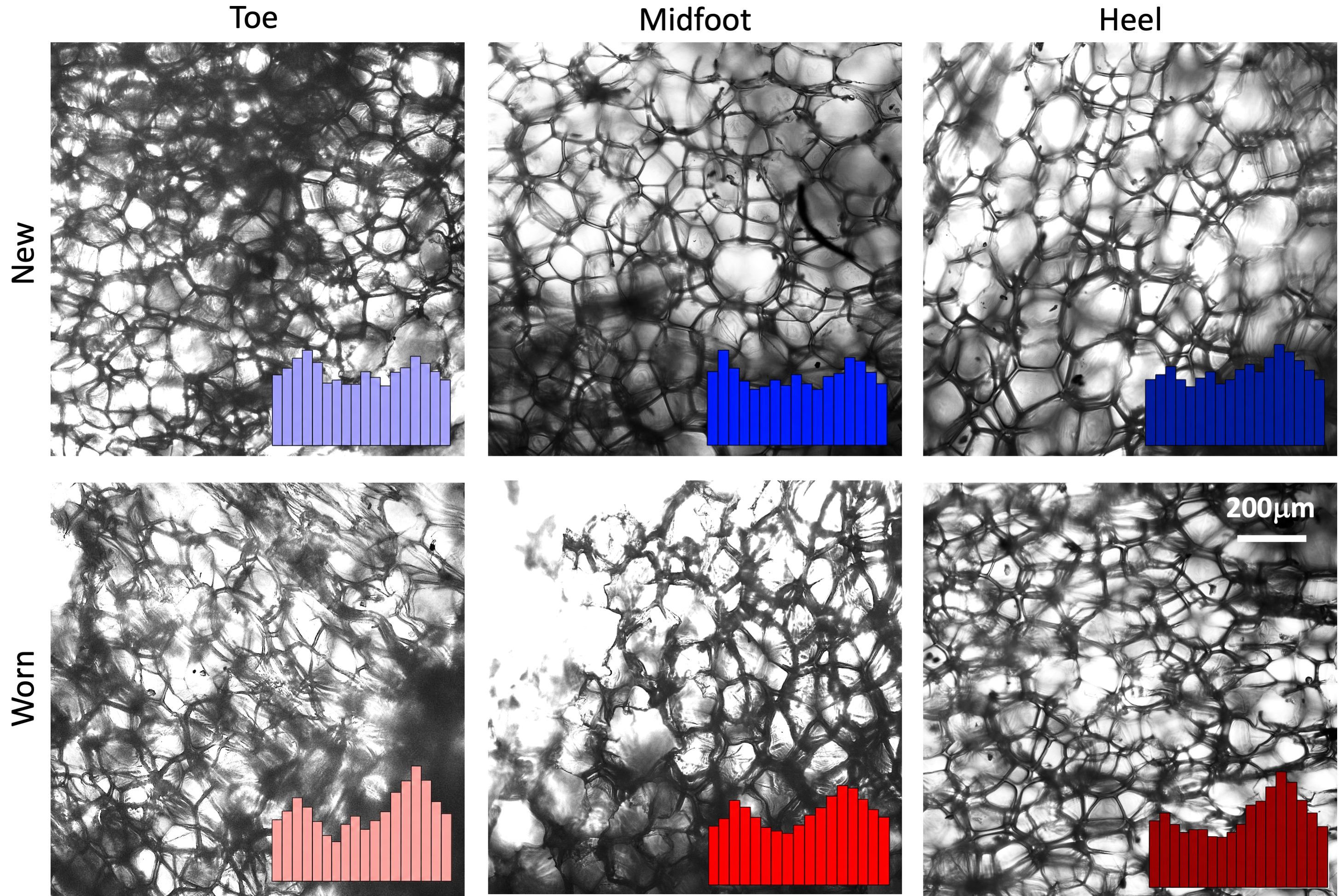}
\caption{{\sffamily{\bfseries{Microstructure and pore-wall orientation of
the new and worn shoes.}}} Representative bright-field microscopy images
from the toe, midfoot, and heel regions of the new and worn shoes.
Across all six images, the characteristic pore length scale is
$128\pm18\,\mu$m. Histograms show the normalized in-plane pore-wall
orientation distributions in $10^\circ$ intervals from $0^\circ$ to
$180^\circ$. The second-order orientation parameter ranges from
$0.055$ to $0.129$ for the new samples and from $0.127$ to
$0.234$ for the worn samples. Together with the absence of a consistent
preferred direction, these low values support an approximately isotropic
cellular microstructure. Scale bar: $200\,\mu$m.}
\label{fig:microscopy}
\end{figure*}
\subsection{Statistics}\label{sec:methods_stats}
Next, we compare the strain energy at maximum deformation and energy return between new and worn samples, across samples from different regions, and across different loading modes. By comparing these scalar quantities, we increase the interpretability of our results while capturing key characteristics of both the elastic and inelastic properties of the foams. To compare two groups of scalars we use a Welch's unequal-variance $t$-test \cite{welch_generalization_1947}. In addition, to compute a two-sided 95\% confidence interval for the difference between two group means we use Welch's unequal-variance $t$-interval. 

Our first question is whether the foam's \textit{mechanical properties change after 300 miles of use}. We perform $t$-tests to compare the energy return and strain energy at maximum deformation between the new and worn groups for each combination of loading mode and region. This results in a total of 18 tests and 18 associated confidence intervals. 
Our second question is whether the foam's \textit{mechanical properties display regional variations}. We perform $t$-tests to compare the energy return and strain energy at maximum deformation for all pairwise comparisons between regions for each loading mode. We now treat all samples from a given region, new or worn, as a single group. This results in 18 more tests and 18 more associated confidence intervals. 
With a total of 36 tests, 
we use the Holm-Bonferroni \cite{holm_simple_1979} method to correct for multiple comparisons. We sort the $p$ values from lowest to highest, 
sequentially compare each $p_k$ with $\alpha/(n+1-k)$, and reject the corresponding null hypotheses until the first $p_k$ fails to meet this criterion, where $\alpha = 0.05$ is our significance threshold, $n = 36$ is the number of tests, and $p_k$ is the $k$-th smallest $p$ value. 
\section{Results}
\subsection{Microstructure}
We successfully collected microscopic images of both shoes from all three regions.
Figure \ref{fig:microscopy} shows the cellular microstructure of the foam
in the toe, midfoot, and heel regions of the new and worn shoes. 
The microscopy images are consistent with a predominantly \textit{closed-cell morphology}, with individual pores largely separated by continuous cell walls. Across all six images, the characteristic pore length scale is
$128\pm18\,\mu$m. At this length scale, the tension and compression
specimens contain on the order of $10^6$ and $10^5$ pores.
The large number of pores relative to the specimen dimensions supports
our treatment of the foam as a {\it{continuum}}.
The normalized pore-wall orientation distributions in Figure
\ref{fig:microscopy} show no consistent preferred orientation.
The new samples have low second-order orientation parameters from
$A_2=0.055$ to $0.129$, while the worn samples have slightly higher
values from $A_2=0.127$ to $0.234$. The preferred orientations vary
across the toe, midfoot, and heel and do not define a consistent material
direction. These observations support our treatment of the foam as
approximately {\it{isotropic}}. We also observe no systematic difference
in characteristic pore length scale between regions or between the new
and worn shoes.
\subsection{Stress vs. stretch}
\begin{figure*}[htbp]
\centering
{\hspace*{1.5cm}\includegraphics[width=0.8\linewidth]{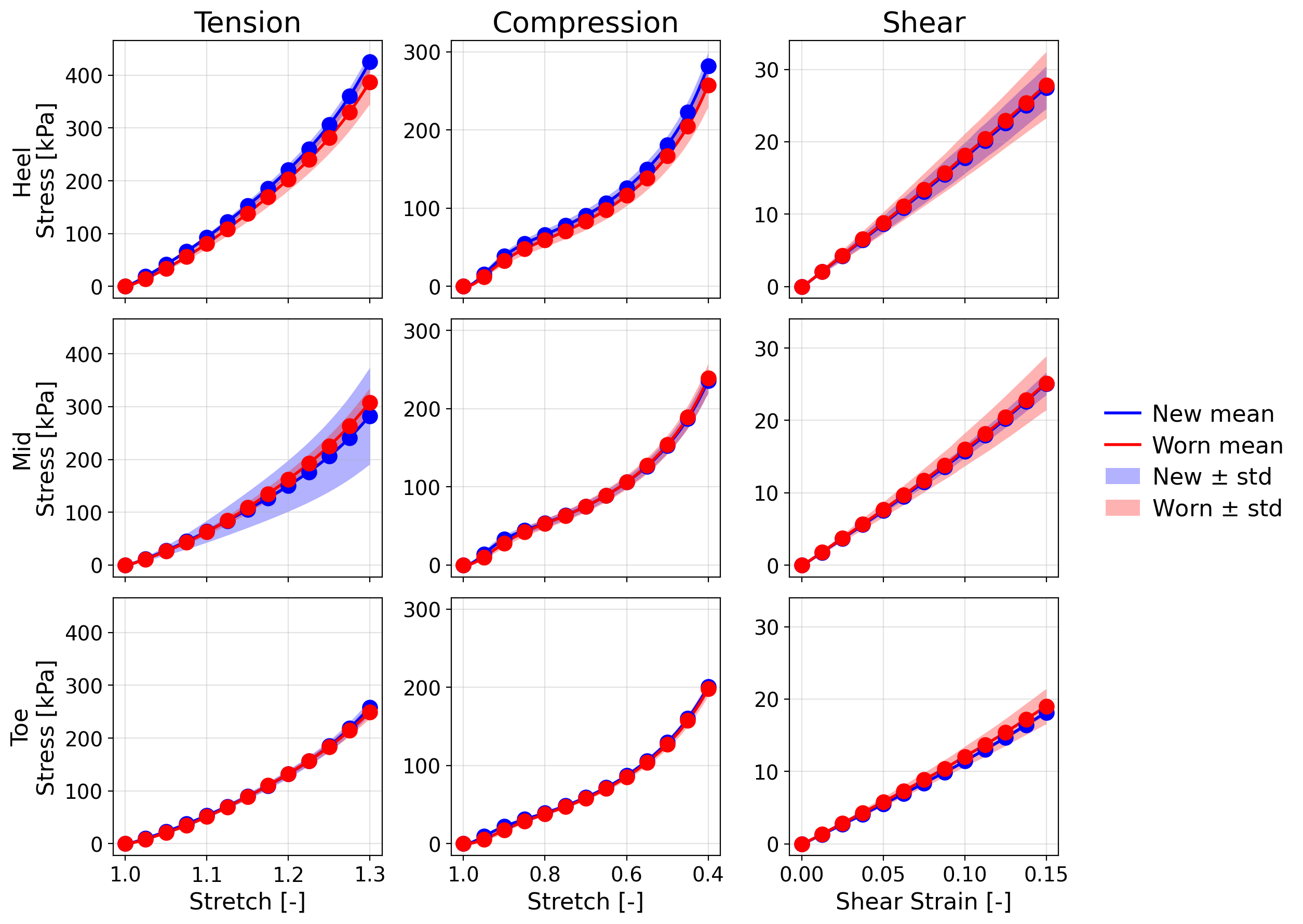}}
\caption{{\sffamily{\bfseries{Stress--stretch response in tension, compression, and shear.}}} Mean stress as a function of stretch for samples from the toe, midfoot, and heel of the new and worn shoes. Shading indicates $\pm$ one standard deviation. Across all three loading modes, the heel displays the stiffest response and the toe the softest, while the characteristic stress--stretch response remains similar between the new and worn shoes.} 
    \label{fig:raw_stress}
\end{figure*}

For each of the three loading modes, tension, compression, and shear, we successfully tested five samples for each of six groups, new toe, new midfoot, new heel, worn toe, worn midfoot, worn heel. 
We plot the means and standard deviations of the stress measurements in Figure \ref{fig:raw_stress} and report them numerically in Tables \ref{tab:raw_data_new_toe} to \ref{tab:raw_data_worn_heel}. 
First and foremost, the nine curves of the new and worn samples lie almost on top of one another, with almost no visual differences. 
While there are small differences in the mean stress between new and worn samples, there is no clear trend across all three regions and loading modes. 
Across all three loading modes and for both new and worn samples, the heel samples have the highest stress (Figure \ref{fig:raw_stress}, top row) and the toe samples have the lowest stress (Figure \ref{fig:raw_stress}, bottom row). 
\subsection{Lateral stretch vs. axial stretch}
\begin{figure*}[t]
    \centering
{\hspace*{1.5cm}\includegraphics[width=0.64\linewidth]{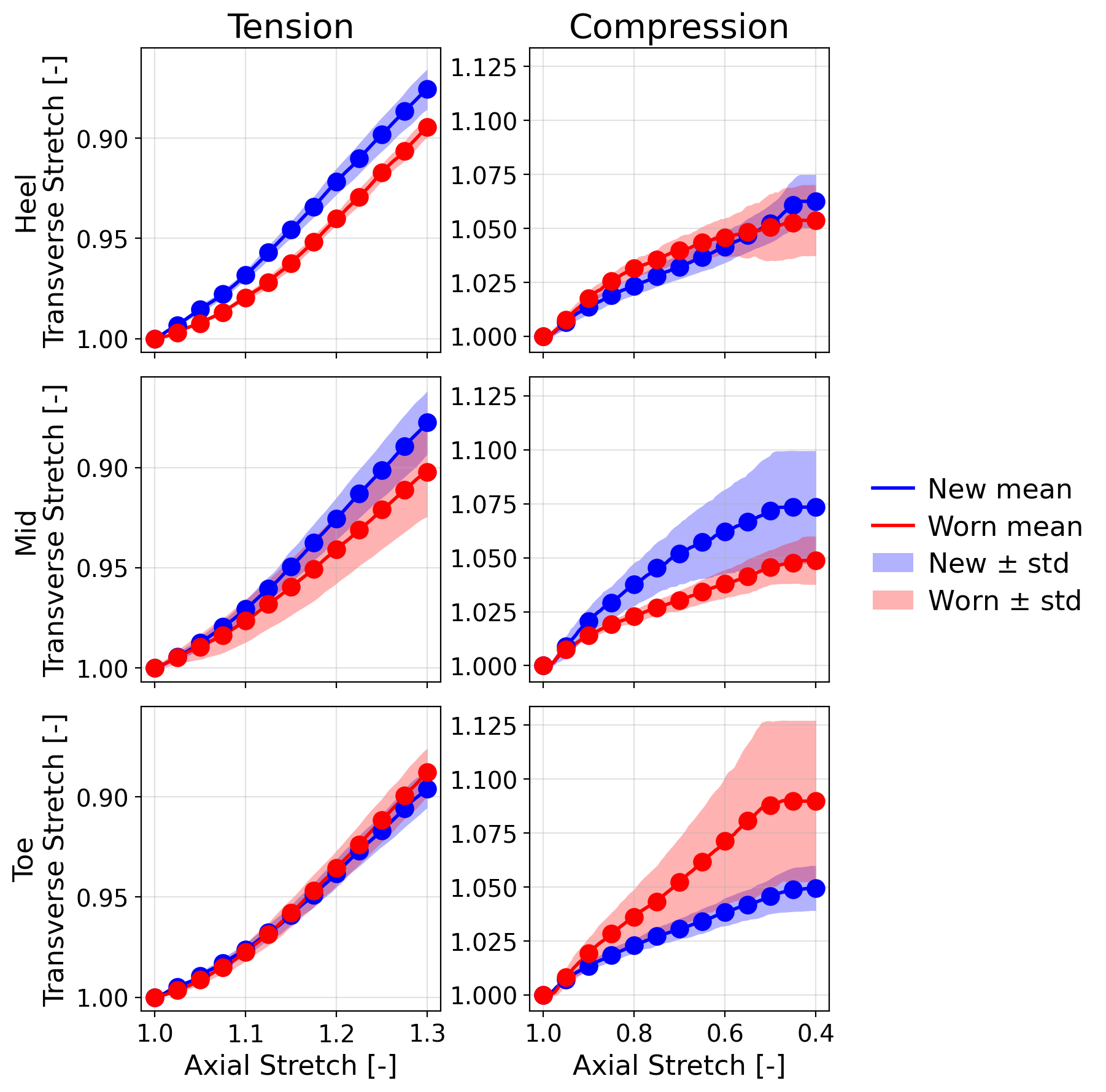}}
\caption{{\sffamily{\bfseries{Transverse stretch in tension and compression.}}} Mean transverse stretch $\lambda_2$ as a function of axial stretch $\lambda_1$ for samples from the toe, midfoot, and heel of the new and worn shoes. Shading indicates $\pm$ one standard deviation. Across all samples, the Poisson's ratio is $\nu_{\rm ten}=0.34\pm0.06$ in tension and $\nu_{\rm com}=0.14\pm0.05$ in compression (mean $\pm$ standard deviation). This pronounced tension--compression asymmetry is statistically significant ($p<0.001$).}    
\label{fig:trans_stretch}
\end{figure*}

For each uniaxial tension and compression experiment, we successfully measured the transverse stretch from a custom video processing pipeline. 
\begin{figure*}
    \centering
    \includegraphics[width=\linewidth]{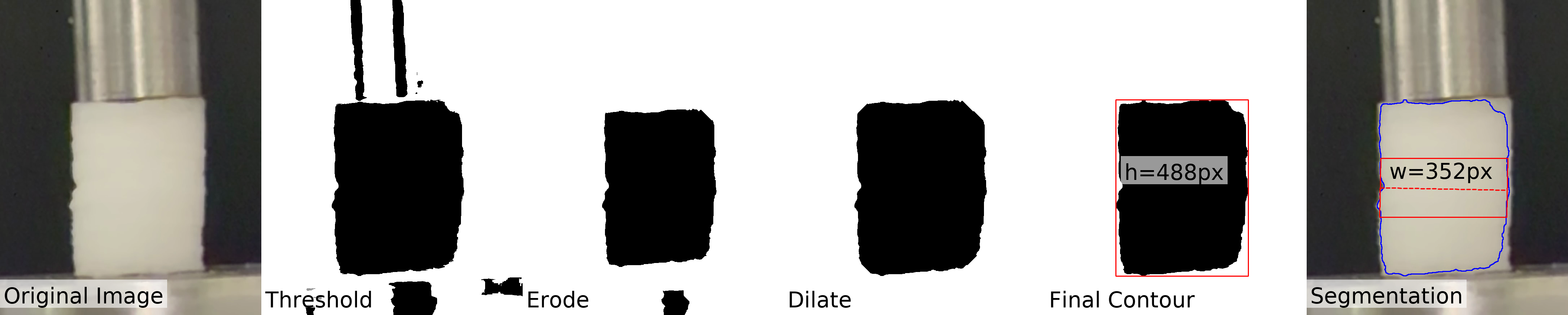}
    \caption{{\sffamily{\bfseries{Video-processing pipeline to assess sample height and width.}}}
    Representative frame from a compression experiment on a toe-region sample from the new shoe at $t$=1\,s with the sequence of processing steps from left to right. Black pixels indicate a mask value of one, white pixels indicate a value of zero. The penultimate panel shows the red bounding box of the final contour, which defines the sample height $h(t)$. The final panel shows the blue final contour and the red parallelogram, which defines sample width $w(t)$.}
    \label{fig:video_frame}
\end{figure*}
Figure \ref{fig:video_frame} shows an example from one of these videos from which we measure the specimen height $h(t)$ and width $w(t)$. 
Both measurements are robust, even in the small-stretch regime. 
We plot the means and standard deviations of the lateral stretch measurements in Figure \ref{fig:trans_stretch} and report them numerically in Tables \ref{tab:raw_data_new_toe} to \ref{tab:raw_data_worn_heel}. Although there is a notable variability in the lateral stretch measurements, there is a clear trend from which we can meaningfully estimate the Poisson's ratio. 
\subsection{Elastic modulus and energy return} 
\begin{figure*}[t]
    \centering
{\hspace*{1.0cm}\includegraphics[width=0.9\linewidth]{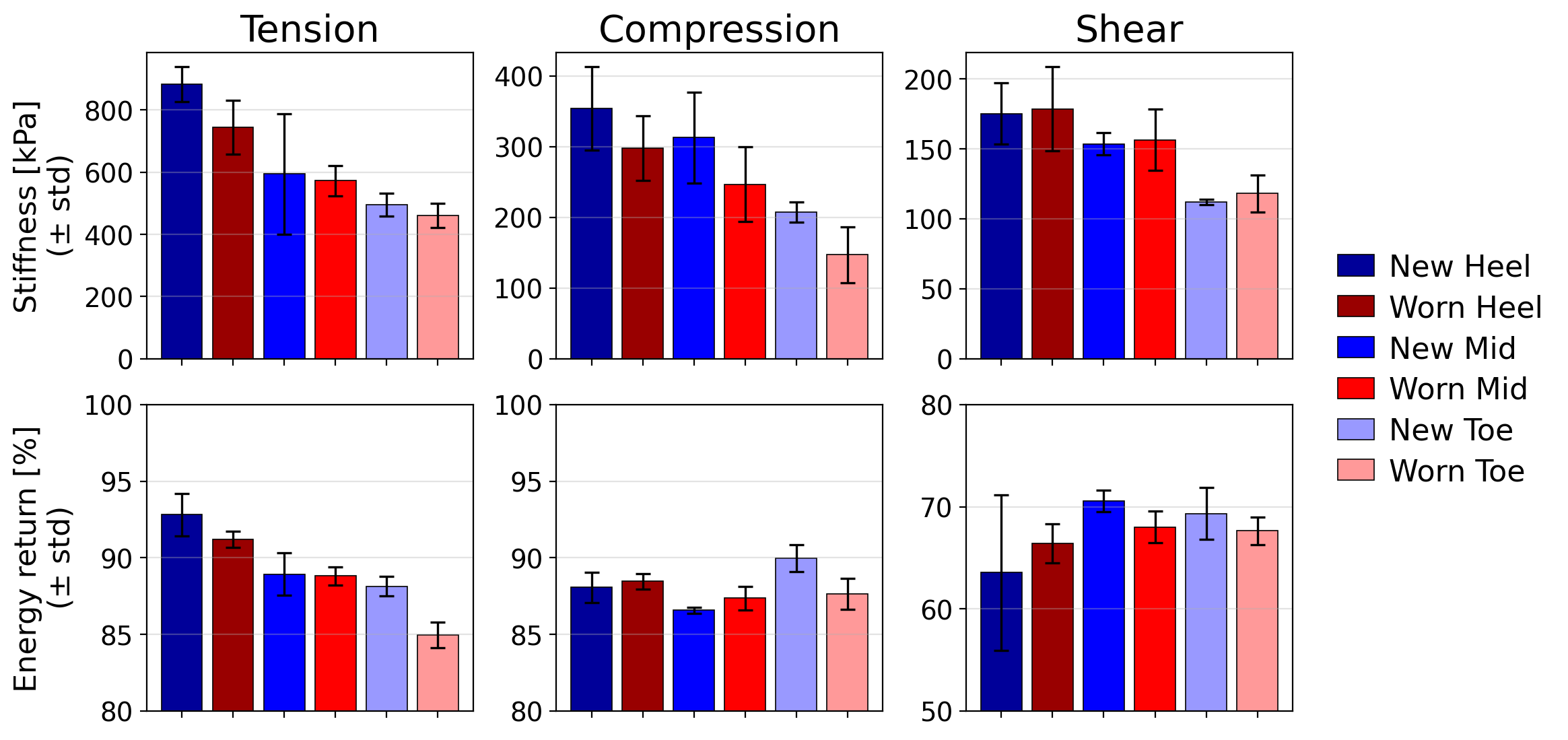}}
\caption{{\sffamily{\bfseries{Elastic modulus and energy return in tension, compression, and shear.}}} The bar height shows the sample mean and the whiskers show $\pm$ one standard deviation for $n=5$ samples. The elastic modulus decreases consistently from heel to midfoot to toe in both the new and worn shoes. Energy return is higher in tension and compression than in shear and shows less systematic variation across regions.}
    \label{fig:scalar_bar_plot}
\end{figure*}

Based on our stress and lateral stretch measurements, we compute the elastic modulus, strain energy at maximum deformation, Poisson's ratio, and energy return for each sample. We then plot the means and standard deviations of the elastic modulus and the energy return for each foam type in Figure \ref{fig:scalar_bar_plot} and report energy return, elastic modulus, and Poisson's ratio in Tables \ref{tab:raw_data_new_toe} to \ref{tab:raw_data_worn_heel}. 
Across all three loading modes, the elastic modulus shows a clear regional variation. The heel is consistently the stiffest region and the toe the least stiff, with the midfoot between the two. This trend is present in both the new and worn shoes and is most pronounced in tension. In contrast, the differences between the new and worn shoes are smaller and do not show a consistent trend across regions or loading modes. The energy return is high in tension and compression, ranging from approximately 85\% to 93\%, and lower in shear, ranging from approximately 64\% to 71\%. Within each loading mode, the energy return varies less systematically across regions and between the new and worn shoes.
To assess the statistical significance of the differences in Figure \ref{fig:scalar_bar_plot}, we perform Welch's $t$-tests and summarize the comparisons of the new and worn shoes in Table \ref{tab:scalar_new_worn} and of the three different regions in Table \ref{tab:scalar_region}.
Of the 36 $t$-tests, ten show statistically significant differences after correcting for multiple comparisons using the Holm-Bonferroni correction with a significance threshold of $\alpha = 0.05$.  
\subsection{Differences between new and worn samples}  
\begin{table*}[t]
    \caption{{\sffamily{\bfseries{Mechanical characteristics of new and worn samples.}}} Differences in strain energy at maximum deformation $\psi$ and energy return $\eta$ between new and worn samples. Values are Welch's $t$-test $p$ values and 95\% confidence intervals CI. Strain-energy differences are reported in kPa and relative to the new samples; energy-return differences are reported in percentage points. Shading indicates statistical significance after Holm-Bonferroni correction at $\alpha=0.05$.}
    \label{tab:scalar_new_worn}
    \centering
    \newcommand{\tablesize}{\footnotesize}
    \setlength{\tabcolsep}{3pt}
    \fontsize{10pt}{11pt}\selectfont
    \begin{tabular}{|c||ccc|ccc|}
\hline
 \makecell{test}   & \makecell{$\psi_{\rm ten}$ \\ {[kPa]}}          & \makecell{$\psi_{\rm com}$ \\ {[kPa]}}             & \makecell{$\psi_{\rm shr}$ \\ {[kPa]}}             & \makecell{$\eta_{\rm ten}$ \\ {[\%]}}   & \makecell{$\eta_{\rm com}$ \\ {[\%]}}   & \makecell{$\eta_{\rm shr}$ \\ {[\%]}}   \\
\hline \hline
 $p$\,(toe, worn vs. new) & $\num{0.69}$                                & $\num{0.31}$                                   & $\num{0.45}$                                   & \cellcolor{gray!20}$0.00037$ & $\num{0.0085}$                      & $\num{0.28}$                        \\
 CI\,(toe, worn - new)& \makecell{$[-3.2,\,2.2]$\\($-11\%$, $7\%$)} & \makecell{$[-4.2,\,1.6]$\\($-9.8\%$, $3.6\%$)} & \makecell{$[-0.15,\,0.29]$\\($-12\%$, $22\%$)} & \makecell{$[-4.4,\,-2]$\\}          & \makecell{$[-3.9,\,-0.8]$\\}        & \makecell{$[-5.2,\,1.8]$\\}         \\ \hline
 $p$\,(midfoot, worn vs. new)  & $\num{0.71}$                                & $\num{0.92}$                                   & $\num{0.87}$                                   & $\num{0.88}$                        & $\num{0.11}$                        & $\num{0.028}$                       \\
 CI\,(midfoot, worn - new)      & \makecell{$[-13,\,17]$\\($-38\%$, $51\%$)}  & \makecell{$[-7.4,\,6.8]$\\($-14\%$, $13\%$)}   & \makecell{$[-0.33,\,0.38]$\\($-18\%$, $21\%$)} & \makecell{$[-2,\,1.8]$\\}           & \makecell{$[-0.3,\,1.8]$\\}         & \makecell{$[-4.7,\,-0.4]$\\}        \\ \hline
 $p$\,(heel, worn vs. new)          & $\num{0.14}$                                & $\num{0.23}$                                   & $\num{0.88}$                                   & $\num{0.078}$                       & $\num{0.50}$                        & $\num{0.50}$                        \\
 CI\,(heel, worn - new)         & \makecell{$[-12,\,2]$\\($-23\%$, $4\%$)}    & \makecell{$[-15,\,4]$\\($-24\%$, $7\%$)}       & \makecell{$[-0.45,\,0.52]$\\($-23\%$, $26\%$)} & \makecell{$[-3.5,\,0.3]$\\}         & \makecell{$[-1,\,1.8]$\\}           & \makecell{$[-8,\,13]$\\}            \\
\hline
\end{tabular}
\vspace*{0.3cm}
\caption{{\sffamily{\bfseries{Mechanical characteristics across shoe regions.}}} Differences in strain energy $\psi$ at maximum deformation and energy return $\eta$ between the toe, midfoot, and heel. Values are Welch's $t$-test $p$ values  and 95\% confidence intervals CI. Strain-energy differences are reported in kPa and relative to the second region; energy-return differences are reported in percentage points. Shading indicates statistical significance after Holm-Bonferroni correction at $\alpha=0.05$.}    
    \label{tab:scalar_region}
    \centering
    \setlength{\tabcolsep}{3pt}
    \begin{tabular}{|c||ccc|ccc|}
\hline
 \makecell{test}                & \makecell{$\psi_{\rm ten}$ \\ {[kPa]}}           & \makecell{$\psi_{\rm com}$ \\ {[kPa]}}           & \makecell{$\psi_{\rm shr}$ \\ {[kPa]}}               & \makecell{$\eta_{\rm ten}$ \\ {[\%]}}   & \makecell{$\eta_{\rm com}$ \\ {[\%]}}   & \makecell{$\eta_{\rm shr}$ \\ {[\%]}}   \\
\hline \hline
 \makecell{$p$\,(toe vs. midfoot)}  & $\num{0.083}$                                & \cellcolor{gray!20}$\num{2e-05}$            & \cellcolor{gray!20}$\num{2e-05}$                & $\num{0.0040}$                      & $\num{0.0056}$                      & $\num{0.41}$                        \\
 \makecell{CI\,(toe - midfoot)}  & \makecell{$[-12,\,0.9]$\\($-33\%$, $2\%$)}   & \makecell{$[-14,\,-7]$\\($-26\%$, $-13\%$)}  & \makecell{$[-0.63,\,-0.3]$\\($-35\%$, $-17\%$)}  & \makecell{$[-3.8,\,-0.9]$\\}        & \makecell{$[0.6,\,3]$\\}            & \makecell{$[-2.8,\,1.2]$\\}         \\ \hline
 \makecell{$p$\,(toe vs. heel)}     & \cellcolor{gray!20}$\num{2e-07}$            & \cellcolor{gray!20}$\num{8e-06}$            & \cellcolor{gray!20}$\num{3e-05}$                & \cellcolor{gray!20}$\num{1e-06}$   & $\num{0.36}$                        & $\num{0.11}$                        \\
 \makecell{CI\,(toe - heel)}     & \makecell{$[-22,\,-15]$\\($-45\%$, $-31\%$)} & \makecell{$[-23,\,-13]$\\($-38\%$, $-22\%$)} & \makecell{$[-0.92,\,-0.46]$\\($-46\%$, $-23\%$)} & \makecell{$[-7,\,-3.9]$\\}          & \makecell{$[-0.7,\,1.7]$\\}         & \makecell{$[-1,\,8]$\\}             \\ \hline
 \makecell{$p$\,(midfoot vs. heel)} & \cellcolor{gray!20}$\num{0.0010}$           & $\num{0.0091}$                               & $\num{0.069}$                                    & \cellcolor{gray!20}$\num{3e-05}$   & \cellcolor{gray!20}$\num{0.0018}$  & $\num{0.055}$                       \\
 \makecell{CI\,(midfoot - heel)} & \makecell{$[-20,\,-6]$\\($-41\%$, $-13\%$)}  & \makecell{$[-13,\,-2]$\\($-22\%$, $-4\%$)}   & \makecell{$[-0.48,\,0.02]$\\($-24\%$, $1\%$)}    & \makecell{$[-4.3,\,-1.9]$\\}        & \makecell{$[-2,\,-0.6]$\\}          & \makecell{$[-0.1,\,8.7]$\\}         \\
\hline
\end{tabular}    
\end{table*}
Remarkably, the mechanical characteristics of the foam remain largely preserved after 300 miles of use. Table \ref{tab:scalar_new_worn} shows no statistically significant difference in strain energy at maximum deformation between the new and worn shoes for any region or loading mode. The differences are also relatively small: for almost all regions and loading modes, the 95\% confidence interval for the change in maximum strain energy lies within $\pm25\%$, and several lie within $\pm10\%$. Moreover, there is no systematic direction of change. The maximum strain energy is greater in the new shoe in five of nine comparisons and greater in the worn shoe in four of nine comparisons. Similarly, energy return remains largely preserved after 300 miles. The only statistically significant difference after correction for multiple comparisons occurs in tension in the toe, where the energy return of the new foam exceeds that of the worn foam by 2.0--4.4 percentage points. For almost all other regions and loading modes, the 95\% confidence intervals remain within $\pm5$ percentage points. Together, these results demonstrate a remarkable retention of the foam's mechanical characteristics after 300 miles of real-world use.
\subsection{Differences across regions}  
In contrast to the small differences between new and worn samples, the foam displays pronounced regional variations in its mechanical response. Table \ref{tab:scalar_region} shows that the strain energy at maximum deformation decreases systematically from heel to midfoot to toe across all three loading modes. The heel and toe differ significantly in tension, compression, and shear, with the strain energy in the toe 31--45\% lower in tension, 22--38\% lower in compression, and 23--46\% lower in shear. The midfoot generally lies between these two extremes, with significant differences from the heel in tension and from the toe in compression and shear. Because this spatial gradient is present across both new and worn samples, it represents a persistent regional characteristic of the foam rather than a progressive consequence of use. Energy return shows smaller regional variations, generally on the order of only a few percentage points. Together, these results reveal a pronounced heel-to-toe gradient in the foam's capacity to store mechanical energy that remains preserved after 300 miles of use.

\subsection{Poisson's ratio} 
Figure \ref{fig:trans_stretch} reveals a pronounced difference in the transverse response between tension and compression. Across all samples, the Poisson's ratio is $\nu_{\rm ten}=0.34\pm0.06$ in tension and $\nu_{\rm com}=0.14\pm0.05$ in compression, reported as mean $\pm$ standard deviation. The difference between tension and compression is highly significant ($p < 10^{-10}$), which reflects a pronounced tension--compression asymmetry in the transverse deformation of the foam.
\section{Discussion}
\noindent{\sffamily{\bfseries{Mechanical properties persist, even after 300 miles.}}}
Across tension, compression, and shear, the stress--stretch curves of
the new and worn foams remain strikingly similar. We find no statistically
significant difference in strain energy at maximum deformation between the
new and worn shoes for any region or loading mode. Energy return is similarly
preserved, with only one statistically significant difference after correction
for multiple comparisons: in tension in the toe, the energy return of the new
foam exceeds that of the worn foam by 2.0-4.4 percentage points. Interestingly,
the differences between new and worn samples do not show a systematic
direction across regions or loading modes. These findings indicate that the
foam retains not only the magnitude of its elastic response, but also its
characteristic nonlinear and dissipative behavior after prolonged use: Within
the limits of this study, \textit{300 miles of real-world running produce remarkably little change} in the material-level mechanical response of the FF
LEAP\textsuperscript{TM} foam. \\[6.pt]
%
\noindent{\sffamily{\bfseries{Mechanical properties vary strongly across regions.}}}
Across all three loading modes, the strain energy at maximum deformation
decreases from heel to midfoot to toe. The toe stores 31--45\% less energy
than the heel in tension, 22--38\% less in compression, and 23--46\% less
in shear. Importantly, this \textit{heel-to-toe gradient} 
is present in both the new and worn shoes. It therefore represents a persistent spatial characteristic
of the foam rather than a progressive consequence of wear. This gradient may arise from spatial variations introduced
during manufacturing, from intentional regional tuning of the midsole, or
from material characteristics that are not resolved by our microscopy.
As a natural consequence, a single material sample cannot fully characterize the mechanical
response of the foam throughout the shoe. This observation is particularly
important for experimental characterization and computational models that
treat a running-shoe midsole as spatially homogeneous.\\[6.pt]
\noindent{\sffamily{\bfseries{Mechanical response depends strongly on loading mode.}}}
Energy return is high in tension and compression, approximately 85--93\%,
but substantially lower in shear, approximately 64--71\%. The transverse
response reveals an additional asymmetry. The mean Poisson's ratio is
$\nu_{\rm ten}=0.34\pm0.06$ in tension and
$\nu_{\rm com}=0.14\pm0.05$ in compression. This pronounced
tension--compression asymmetry is consistent with the mechanics of cellular
solids, whose pore structure can deform differently as cell walls stretch,
bend, buckle, and collapse under different loading modes. 
The predominantly closed-cell morphology observed in our microscopy may contribute to this asymmetry: compression promotes cell-wall bending and buckling together with changes in cell geometry and pressure within the enclosed pores, while tension primarily stretches and reorients the cell walls.
A single elastic
modulus or Poisson's ratio is therefore insufficient to describe the
mechanical response of this ultra-low-density foam. The complete axial and
transverse deformation data reported here provide a richer basis for
constitutive models that account for nonlinear response, compressibility,
and tension--compression asymmetry.\\[6.pt]
\noindent{\sffamily{\bfseries{Microstructure supports a continuum description.}}}
Across all six microscopy images, the characteristic pore length scale is
$128\pm18\,\mu$m. At this scale, the tension and compression specimens
contain on the order of $10^6$ and $10^5$ pores, which provides a clear separation between the pore and specimen length scales.
The orientation analysis yields low second-order orientation parameters
of $0.055$-$0.129$ in the new and
$0.127$-$0.234$ in the worn samples,
where zero corresponds to a uniform orientation distribution and one to perfect alignment. Although the worn samples show
somewhat larger orientation parameters, the preferred directions differ
between the toe, midfoot, and heel and do not define a consistent material
direction or trend. 
Taken together, these observations support an approximately {\it{isotropic continuum}} description at the specimen scale. 
Since we do not observe any systematic variations in
characteristic pore length scale between regions or between the new and worn
shoes, we conclude that pore size alone cannot explain the pronounced regional differences in the mechanical response.
Other microstructural features, including local density, cell-wall thickness,
cell shape, or material composition, may contribute to the observed
heel-to-toe gradient.\\[6.pt]
\noindent{\sffamily{\bfseries{The study has a few limitations.}}}
Most importantly, we compare one new shoe with one shoe worn by a single
runner for 300 miles. Running speed, body mass, foot-strike pattern, terrain,
temperature, and training history can all influence the mechanical loading
experienced by a midsole. Studies across multiple runners and shoes are
therefore necessary to quantify variability in the rate and spatial pattern
of material degradation. In addition, our experiments characterize isolated
foam samples rather than the complete shoe. They do not directly quantify
running economy, biomechanical performance, injury risk, or degradation of
other shoe components such as the carbon-fiber plate and upper. Finally, our
microscopy provides two-dimensional views of a three-dimensional cellular
material and cannot fully resolve pore connectivity. The orientation analysis supports approximate in-plane isotropy, but does not exclude possible out-of-plane anisotropy.\\[6.pt]
\noindent{\sffamily{\bfseries{The foam exhibits remarkable mechanical resilience.}}}
After 300 miles of real-world use, the foam retains its nonlinear stress--deformation response, strain-energy capacity, energy-return characteristics, pronounced tension--compression asymmetry, and persistent mechanical heel-to-toe gradient. If this material-level resilience extends across runners and shoes, our findings challenge the common practice of replacing high-performance racing shoes after approximately 300 miles because of concerns about midsole foam degradation. From the perspective of the foam alone, our results provide \textit{no mechanical evidence that replacement is necessary} at this mileage. Other shoe components, including the carbon-fiber plate and upper, may degrade more rapidly and ultimately determine the lifetime of the shoe. These measurements provide a quantitative foundation for constitutive models of high-performance running shoes and motivate future studies that connect material-level durability to whole-shoe mechanics and running performance.
\section{Conclusion}
\label{sec:conclusion}
High-performance racing shoes rely on ultra-low-density elastomeric foams
that undergo large, repeated deformations during running. Here, we show that
the characteristic mechanical response of FF LEAP\textsuperscript{TM} foam
remains remarkably resilient after 300 miles of real-world use. Across
tension, compression, and shear, the worn foam largely retains its nonlinear
stress--deformation response, strain-energy capacity, and energy-return
characteristics. At the same time, the foam exhibits a pronounced mechanical
heel-to-toe gradient and a strongly mode-dependent response, including a
pronounced tension--compression asymmetry in Poisson's ratio. Microscopy
reveals a characteristic pore length scale of $128\pm18\,\mu$m and supports
an approximately isotropic continuum description of the foam.
Together, these findings suggest that the intrinsic mechanical characteristics
of this ultra-low-density racing-shoe foam remain largely preserved after
300 miles of use. If this resilience extends across runners and shoes, midsole foam degradation may not justify replacing a racing shoe after 300 miles. Other shoe components may ultimately determine the lifetime of the shoe.
Beyond durability, the regional and loading-mode-dependent properties
identified here provide a quantitative foundation for constitutive models
of high-performance running shoes and for future studies that connect
material-level mechanics to whole-shoe performance.
\section*{Appendix: Stress and transverse stretch data}
\label{appendix:raw_data}
In Tables \ref{tab:raw_data_new_toe} to \ref{tab:raw_data_worn_heel}, we report processed stress and transverse stretch measurements as means $\pm$ standard deviations, resampled at 13 equidistant stretch or shear strain states, for each region of the new and worn shoes. We also report the means and standard deviations of the elastic modulus, Poisson's ratio, and energy return for each loading mode. 
\subsection*{Acknowledgments}
We thank Benjamin Alfred Jones for helping with the collection of microscopy images. 
Jeremy McCulloch acknowledges 
the Wu Tsai Human Performance Alliance Digital Athlete Fellowship; 
Ellen Kuhl also acknowledges
the Wu Tsai Human Performance Alliance,
the NSF CMMI grant 2320933, and 
the ERC Advanced Grant 101141626. 
\printcredits
\begin{table*}[p]
\caption{{\sffamily{\bfseries{
Stress, transverse stretch, and mechanical properties of the toe region
of the new shoe.}}}
Values report mean $\pm$ standard deviation for $n=5$ samples in tension,
compression, and shear, including the elastic modulus, Poisson's ratio, and energy return.}
\label{tab:raw_data_new_toe}
    \centering
    \newcommand{\tablesize}{\footnotesize}
        \fontsize{10pt}{11pt}\selectfont
    \setlength{\tabcolsep}{3pt}    
\begin{tabular}{|ccc||ccc||cc|}
\hline
  \multicolumn{3}{|c||}{\sffamily{\bfseries{uniaxial tension}}}
& \multicolumn{3}{c||} {\sffamily{\bfseries{uniaxial compression}}}
& \multicolumn{2}{c|}  {\sffamily{\bfseries{simple shear}}} \\
  \multicolumn{3}{|c||}{$n=5$}
& \multicolumn{3}{c||}{$n=5$}
& \multicolumn{2}{c|}{$n=5$} \\ \hline
$\lambda_1$ & $\lambda_2$ & $P_{11}$ & $\lambda_1$ & $\lambda_2$ & $|P_{11}|$ & $\gamma$ & $P_{12}$ \\
\,[-] & [-] & [kPa] & [-] & [-] & [kPa] & [-] & [kPa] \\
\hline \hline
1.000 & 1.000\hspace{0.5em}$\pm$ 0.000 & \phantom{0}\phantom{0}0.00\hspace{0.5em}$\pm$ \phantom{0}0.00 & 1.000 & 1.000\hspace{0.5em}$\pm$ 0.000 & \phantom{0}\phantom{0}0.00\hspace{0.5em}$\pm$ \phantom{0}0.00 & 0.000 & \phantom{0}0.00\hspace{0.5em}$\pm$ 0.00 \\ \hline
1.025 & 0.995\hspace{0.5em}$\pm$ 0.001 & \phantom{0}\phantom{0}9.90\hspace{0.5em}$\pm$ \phantom{0}1.02 & 0.950 & 1.007\hspace{0.5em}$\pm$ 0.001 & \phantom{0}\phantom{0}9.33\hspace{0.5em}$\pm$ \phantom{0}1.03 & 0.012 & \phantom{0}1.30\hspace{0.5em}$\pm$ 0.03 \\
1.050 & 0.990\hspace{0.5em}$\pm$ 0.002 & \phantom{0}22.94\hspace{0.5em}$\pm$ \phantom{0}1.86 & 0.900 & 1.013\hspace{0.5em}$\pm$ 0.002 & \phantom{0}22.02\hspace{0.5em}$\pm$ \phantom{0}1.06 & 0.025 & \phantom{0}2.68\hspace{0.5em}$\pm$ 0.05 \\
1.075 & 0.983\hspace{0.5em}$\pm$ 0.002 & \phantom{0}37.36\hspace{0.5em}$\pm$ \phantom{0}2.70 & 0.850 & 1.019\hspace{0.5em}$\pm$ 0.003 & \phantom{0}31.18\hspace{0.5em}$\pm$ \phantom{0}1.21 & 0.037 & \phantom{0}4.08\hspace{0.5em}$\pm$ 0.08 \\ \hline
1.100 & 0.976\hspace{0.5em}$\pm$ 0.003 & \phantom{0}53.26\hspace{0.5em}$\pm$ \phantom{0}3.67 & 0.800 & 1.023\hspace{0.5em}$\pm$ 0.003 & \phantom{0}39.40\hspace{0.5em}$\pm$ \phantom{0}1.43 & 0.050 & \phantom{0}5.49\hspace{0.5em}$\pm$ 0.10 \\ \hline
1.125 & 0.968\hspace{0.5em}$\pm$ 0.004 & \phantom{0}70.65\hspace{0.5em}$\pm$ \phantom{0}4.73 & 0.750 & 1.027\hspace{0.5em}$\pm$ 0.004 & \phantom{0}48.51\hspace{0.5em}$\pm$ \phantom{0}1.70 & 0.062 & \phantom{0}6.92\hspace{0.5em}$\pm$ 0.13 \\
1.150 & 0.959\hspace{0.5em}$\pm$ 0.005 & \phantom{0}89.60\hspace{0.5em}$\pm$ \phantom{0}5.84 & 0.700 & 1.031\hspace{0.5em}$\pm$ 0.005 & \phantom{0}59.23\hspace{0.5em}$\pm$ \phantom{0}2.06 & 0.075 & \phantom{0}8.39\hspace{0.5em}$\pm$ 0.15 \\
1.175 & 0.949\hspace{0.5em}$\pm$ 0.006 & 110.01\hspace{0.5em}$\pm$ \phantom{0}6.89 & 0.650 & 1.034\hspace{0.5em}$\pm$ 0.006 & \phantom{0}71.88\hspace{0.5em}$\pm$ \phantom{0}2.38 & 0.087 & \phantom{0}9.89\hspace{0.5em}$\pm$ 0.18 \\ \hline
1.200 & 0.938\hspace{0.5em}$\pm$ 0.007 & 132.14\hspace{0.5em}$\pm$ \phantom{0}7.95 & 0.600 & 1.038\hspace{0.5em}$\pm$ 0.006 & \phantom{0}87.21\hspace{0.5em}$\pm$ \phantom{0}2.85 & 0.100 & 11.45\hspace{0.5em}$\pm$ 0.20 \\ \hline
1.225 & 0.927\hspace{0.5em}$\pm$ 0.008 & 156.73\hspace{0.5em}$\pm$ \phantom{0}9.09 & 0.550 & 1.042\hspace{0.5em}$\pm$ 0.007 & 106.08\hspace{0.5em}$\pm$ \phantom{0}3.41 & 0.112 & 13.06\hspace{0.5em}$\pm$ 0.22 \\
1.250 & 0.917\hspace{0.5em}$\pm$ 0.008 & 184.94\hspace{0.5em}$\pm$ 10.44 & 0.500 & 1.046\hspace{0.5em}$\pm$ 0.008 & 129.47\hspace{0.5em}$\pm$ \phantom{0}4.21 & 0.125 & 14.69\hspace{0.5em}$\pm$ 0.25 \\
1.275 & 0.906\hspace{0.5em}$\pm$ 0.009 & 218.47\hspace{0.5em}$\pm$ 12.14 & 0.450 & 1.049\hspace{0.5em}$\pm$ 0.010 & 160.10\hspace{0.5em}$\pm$ \phantom{0}5.30 & 0.137 & 16.43\hspace{0.5em}$\pm$ 0.29 \\ \hline
1.300 & 0.896\hspace{0.5em}$\pm$ 0.010 & 257.69\hspace{0.5em}$\pm$ 14.29 & 0.400 & 1.050\hspace{0.5em}$\pm$ 0.010 & 201.08\hspace{0.5em}$\pm$ \phantom{0}6.77 & 0.150 & 18.15\hspace{0.5em}$\pm$ 0.33 \\
\hline \hline
  \multicolumn{3}{|c||}{\sffamily{\bfseries{tensile modulus}}}
& \multicolumn{3}{c||} {\sffamily{\bfseries{compressive modulus}}}
& \multicolumn{2}{c|}  {\sffamily{\bfseries{shear modulus}}} \\
  \multicolumn{3}{|c||}{$\textsf{E}_{\rm{ten}} = 495.93 \pm 36.44$\,kPa}
& \multicolumn{3}{c||} {$\textsf{E}_{\rm{com}} = 207.30 \pm 14.32$\,kPa}
& \multicolumn{2}{c|}  {$\textsf{G} = 111.92 \pm 2.03$\,kPa} \\
\hline \hline
  \multicolumn{3}{|c||}{\sffamily{\bfseries{tensile Poisson's ratio}}}
& \multicolumn{3}{c||} {\sffamily{\bfseries{compressive Poisson's ratio}}}
& \multicolumn{2}{c|}{} \\
  \multicolumn{3}{|c||}{$\nu_{\rm{ten}} = 0.317 \pm 0.033$}
& \multicolumn{3}{c||} {$\nu_{\rm{com}} = 0.112 \pm 0.017$}
& \multicolumn{2}{c|}{} \\
\hline \hline
  \multicolumn{3}{|c||}{\sffamily{\bfseries{energy return}}}
& \multicolumn{3}{c||} {\sffamily{\bfseries{energy return}}}
& \multicolumn{2}{c|}  {\sffamily{\bfseries{energy return}}} \\
  \multicolumn{3}{|c||}{$\eta_{\rm{ten}}  = 88.1 \pm 0.6 \%$}
& \multicolumn{3}{c||} {$\eta_{\rm{com}}  = 90.0 \pm 0.9\%$}
& \multicolumn{2}{c|}  {$\eta_{\rm{shr}}  = 69.3 \pm 2.6 \%$} \\
\hline
\end{tabular}    
\vspace*{0.5cm}
\caption{{\sffamily{\bfseries{
Stress, transverse stretch, and mechanical properties of the toe region
of the worn shoe.}}}
Values report mean $\pm$ standard deviation for $n=5$ samples in tension,
compression, and shear, including the elastic modulus, Poisson's ratio, and energy return.}
    \label{tab:raw_data_worn_toe}
    \centering
    \setlength{\tabcolsep}{3pt}
            \fontsize{10pt}{11pt}\selectfont
            \begin{tabular}{|ccc||ccc||cc|}
\hline
  \multicolumn{3}{|c||}{\sffamily{\bfseries{uniaxial tension}}}
& \multicolumn{3}{c||} {\sffamily{\bfseries{uniaxial compression}}}
& \multicolumn{2}{c|}  {\sffamily{\bfseries{simple shear}}} \\
  \multicolumn{3}{|c||}{$n=5$}
& \multicolumn{3}{c||}{$n=5$}
& \multicolumn{2}{c|}{$n=5$} \\ \hline
$\lambda_1$ & $\lambda_2$ & $P_{11}$ & $\lambda_1$ & $\lambda_2$ & $|P_{11}|$ & $\gamma$ & $P_{12}$ \\
\,[-] & [-] & [kPa] & [-] & [-] & [kPa] & [-] & [kPa] \\
\hline \hline
1.000 & 1.000\hspace{0.5em}$\pm$ 0.000 & \phantom{0}\phantom{0}0.00\hspace{0.5em}$\pm$ \phantom{0}0.00 & 1.000 & 1.000\hspace{0.5em}$\pm$ 0.000 & \phantom{0}\phantom{0}0.00\hspace{0.5em}$\pm$ \phantom{0}0.00 & 0.000 & \phantom{0}0.00\hspace{0.5em}$\pm$ 0.00 \\ \hline
1.025 & 0.997\hspace{0.5em}$\pm$ 0.001 & \phantom{0}\phantom{0}8.39\hspace{0.5em}$\pm$ \phantom{0}1.47 & 0.950 & 1.008\hspace{0.5em}$\pm$ 0.004 & \phantom{0}\phantom{0}5.61\hspace{0.5em}$\pm$ \phantom{0}2.14 & 0.012 & \phantom{0}1.38\hspace{0.5em}$\pm$ 0.15 \\
1.050 & 0.991\hspace{0.5em}$\pm$ 0.002 & \phantom{0}20.65\hspace{0.5em}$\pm$ \phantom{0}2.38 & 0.900 & 1.019\hspace{0.5em}$\pm$ 0.007 & \phantom{0}17.57\hspace{0.5em}$\pm$ \phantom{0}3.38 & 0.025 & \phantom{0}2.84\hspace{0.5em}$\pm$ 0.30 \\
1.075 & 0.985\hspace{0.5em}$\pm$ 0.003 & \phantom{0}34.71\hspace{0.5em}$\pm$ \phantom{0}2.92 & 0.850 & 1.028\hspace{0.5em}$\pm$ 0.010 & \phantom{0}28.60\hspace{0.5em}$\pm$ \phantom{0}2.37 & 0.037 & \phantom{0}4.31\hspace{0.5em}$\pm$ 0.46 \\ \hline
1.100 & 0.977\hspace{0.5em}$\pm$ 0.004 & \phantom{0}51.02\hspace{0.5em}$\pm$ \phantom{0}3.24 & 0.800 & 1.036\hspace{0.5em}$\pm$ 0.013 & \phantom{0}37.91\hspace{0.5em}$\pm$ \phantom{0}1.97 & 0.050 & \phantom{0}5.80\hspace{0.5em}$\pm$ 0.63 \\ \hline
1.125 & 0.969\hspace{0.5em}$\pm$ 0.006 & \phantom{0}69.27\hspace{0.5em}$\pm$ \phantom{0}3.68 & 0.750 & 1.043\hspace{0.5em}$\pm$ 0.017 & \phantom{0}47.28\hspace{0.5em}$\pm$ \phantom{0}2.17 & 0.062 & \phantom{0}7.32\hspace{0.5em}$\pm$ 0.80 \\
1.150 & 0.958\hspace{0.5em}$\pm$ 0.007 & \phantom{0}89.13\hspace{0.5em}$\pm$ \phantom{0}4.34 & 0.700 & 1.052\hspace{0.5em}$\pm$ 0.020 & \phantom{0}57.90\hspace{0.5em}$\pm$ \phantom{0}2.61 & 0.075 & \phantom{0}8.85\hspace{0.5em}$\pm$ 0.99 \\
1.175 & 0.947\hspace{0.5em}$\pm$ 0.008 & 110.21\hspace{0.5em}$\pm$ \phantom{0}5.23 & 0.650 & 1.062\hspace{0.5em}$\pm$ 0.025 & \phantom{0}70.39\hspace{0.5em}$\pm$ \phantom{0}3.08 & 0.087 & 10.42\hspace{0.5em}$\pm$ 1.19 \\ \hline
1.200 & 0.936\hspace{0.5em}$\pm$ 0.009 & 132.61\hspace{0.5em}$\pm$ \phantom{0}6.24 & 0.600 & 1.071\hspace{0.5em}$\pm$ 0.030 & \phantom{0}85.50\hspace{0.5em}$\pm$ \phantom{0}3.75 & 0.100 & 12.07\hspace{0.5em}$\pm$ 1.40 \\ \hline
1.225 & 0.924\hspace{0.5em}$\pm$ 0.010 & 156.79\hspace{0.5em}$\pm$ \phantom{0}7.40 & 0.550 & 1.081\hspace{0.5em}$\pm$ 0.036 & 104.11\hspace{0.5em}$\pm$ \phantom{0}4.55 & 0.112 & 13.71\hspace{0.5em}$\pm$ 1.64 \\
1.250 & 0.912\hspace{0.5em}$\pm$ 0.011 & 183.63\hspace{0.5em}$\pm$ \phantom{0}8.79 & 0.500 & 1.088\hspace{0.5em}$\pm$ 0.039 & 127.23\hspace{0.5em}$\pm$ \phantom{0}5.58 & 0.125 & 15.43\hspace{0.5em}$\pm$ 1.87 \\
1.275 & 0.899\hspace{0.5em}$\pm$ 0.012 & 214.41\hspace{0.5em}$\pm$ 10.50 & 0.450 & 1.090\hspace{0.5em}$\pm$ 0.037 & 157.44\hspace{0.5em}$\pm$ \phantom{0}7.19 & 0.137 & 17.23\hspace{0.5em}$\pm$ 2.15 \\ \hline
1.300 & 0.888\hspace{0.5em}$\pm$ 0.012 & 248.78\hspace{0.5em}$\pm$ 12.79 & 0.400 & 1.090\hspace{0.5em}$\pm$ 0.037 & 198.48\hspace{0.5em}$\pm$ \phantom{0}9.10 & 0.150 & 18.99\hspace{0.5em}$\pm$ 2.44 \\
\hline \hline
  \multicolumn{3}{|c||}{\sffamily{\bfseries{tensile modulus}}}
& \multicolumn{3}{c||} {\sffamily{\bfseries{compressive modulus}}}
& \multicolumn{2}{c|}  {\sffamily{\bfseries{shear modulus}}} \\
  \multicolumn{3}{|c||}{$\textsf{E}_{\rm{ten}} = 461.29 \pm 38.92$\,kPa}
& \multicolumn{3}{c||} {$\textsf{E}_{\rm{com}} = 147.22 \pm 39.67$\,kPa}
& \multicolumn{2}{c|}  {$\textsf{G} = 118.15 \pm 13.23$\,kPa} \\
\hline \hline
  \multicolumn{3}{|c||}{\sffamily{\bfseries{tensile Poisson's ratio}}}
& \multicolumn{3}{c||} {\sffamily{\bfseries{compressive Poisson's ratio}}}
& \multicolumn{2}{c|}{} \\
  \multicolumn{3}{|c||}{$\nu_{\rm{ten}} = 0.378 \pm 0.053$}
& \multicolumn{3}{c||} {$\nu_{\rm{com}} = 0.184 \pm 0.049$}
& \multicolumn{2}{c|}{} \\
\hline \hline
  \multicolumn{3}{|c||}{\sffamily{\bfseries{energy return}}}
& \multicolumn{3}{c||} {\sffamily{\bfseries{energy return}}}
& \multicolumn{2}{c|}  {\sffamily{\bfseries{energy return}}} \\
  \multicolumn{3}{|c||}{$\eta_{\rm{ten}}  = 85.0 \pm 0.8 \%$}
& \multicolumn{3}{c||} {$\eta_{\rm{com}}  = 87.6 \pm 1.0\%$}
& \multicolumn{2}{c|}  {$\eta_{\rm{shr}}  = 67.6 \pm 1.3 \%$} \\
\hline
\end{tabular}
\end{table*}

\begin{table*}[p]
\caption{{\sffamily{\bfseries{
Stress, transverse stretch, and mechanical properties of the midfoot region
of the new shoe.}}}
Values report mean $\pm$ standard deviation for $n=5$ samples in tension,
compression, and shear, including the elastic modulus, Poisson's ratio, and energy return.}
\label{tab:raw_data_new_mid}
    \centering
    \newcommand{\tablesize}{\footnotesize}
    \setlength{\tabcolsep}{3pt}
    \fontsize{10pt}{11pt}\selectfont
\begin{tabular}{|ccc||ccc||cc|}
\hline
  \multicolumn{3}{|c||}{\sffamily{\bfseries{uniaxial tension}}}
& \multicolumn{3}{c||} {\sffamily{\bfseries{uniaxial compression}}}
& \multicolumn{2}{c|}  {\sffamily{\bfseries{simple shear}}} \\
  \multicolumn{3}{|c||}{$n=5$}
& \multicolumn{3}{c||}{$n=5$}
& \multicolumn{2}{c|}{$n=5$} \\ \hline
$\lambda_1$ & $\lambda_2$ & $P_{11}$ & $\lambda_1$ & $\lambda_2$ & $|P_{11}|$ & $\gamma$ & $P_{12}$ \\
\,[-] & [-] & [kPa] & [-] & [-] & [kPa] & [-] & [kPa] \\
\hline \hline
1.000 & 1.000\hspace{0.5em}$\pm$ 0.000 & \phantom{0}\phantom{0}0.00\hspace{0.5em}$\pm$ \phantom{0}0.00 & 1.000 & 1.000\hspace{0.5em}$\pm$ 0.000 & \phantom{0}\phantom{0}0.00\hspace{0.5em}$\pm$ \phantom{0}0.00 & 0.000 & \phantom{0}0.00\hspace{0.5em}$\pm$ 0.00 \\ \hline
1.025 & 0.994\hspace{0.5em}$\pm$ 0.002 & \phantom{0}11.98\hspace{0.5em}$\pm$ \phantom{0}3.90 & 0.950 & 1.009\hspace{0.5em}$\pm$ 0.005 & \phantom{0}14.13\hspace{0.5em}$\pm$ \phantom{0}3.81 & 0.012 & \phantom{0}1.78\hspace{0.5em}$\pm$ 0.09 \\
1.050 & 0.987\hspace{0.5em}$\pm$ 0.003 & \phantom{0}27.50\hspace{0.5em}$\pm$ \phantom{0}8.94 & 0.900 & 1.020\hspace{0.5em}$\pm$ 0.006 & \phantom{0}33.06\hspace{0.5em}$\pm$ \phantom{0}5.41 & 0.025 & \phantom{0}3.68\hspace{0.5em}$\pm$ 0.18 \\
1.075 & 0.980\hspace{0.5em}$\pm$ 0.003 & \phantom{0}44.76\hspace{0.5em}$\pm$ 14.66 & 0.850 & 1.029\hspace{0.5em}$\pm$ 0.008 & \phantom{0}44.34\hspace{0.5em}$\pm$ \phantom{0}5.11 & 0.037 & \phantom{0}5.59\hspace{0.5em}$\pm$ 0.28 \\ \hline
1.100 & 0.971\hspace{0.5em}$\pm$ 0.005 & \phantom{0}63.65\hspace{0.5em}$\pm$ 20.88 & 0.800 & 1.038\hspace{0.5em}$\pm$ 0.010 & \phantom{0}53.61\hspace{0.5em}$\pm$ \phantom{0}5.31 & 0.050 & \phantom{0}7.52\hspace{0.5em}$\pm$ 0.38 \\ \hline
1.125 & 0.961\hspace{0.5em}$\pm$ 0.006 & \phantom{0}83.79\hspace{0.5em}$\pm$ 27.34 & 0.750 & 1.045\hspace{0.5em}$\pm$ 0.012 & \phantom{0}63.64\hspace{0.5em}$\pm$ \phantom{0}5.67 & 0.062 & \phantom{0}9.49\hspace{0.5em}$\pm$ 0.49 \\
1.150 & 0.949\hspace{0.5em}$\pm$ 0.008 & 104.81\hspace{0.5em}$\pm$ 34.04 & 0.700 & 1.052\hspace{0.5em}$\pm$ 0.014 & \phantom{0}75.28\hspace{0.5em}$\pm$ \phantom{0}6.06 & 0.075 & 11.50\hspace{0.5em}$\pm$ 0.61 \\
1.175 & 0.937\hspace{0.5em}$\pm$ 0.009 & 126.75\hspace{0.5em}$\pm$ 41.07 & 0.650 & 1.057\hspace{0.5em}$\pm$ 0.017 & \phantom{0}89.10\hspace{0.5em}$\pm$ \phantom{0}6.52 & 0.087 & 13.58\hspace{0.5em}$\pm$ 0.73 \\ \hline
1.200 & 0.926\hspace{0.5em}$\pm$ 0.011 & 150.25\hspace{0.5em}$\pm$ 48.66 & 0.600 & 1.062\hspace{0.5em}$\pm$ 0.020 & 105.90\hspace{0.5em}$\pm$ \phantom{0}7.18 & 0.100 & 15.73\hspace{0.5em}$\pm$ 0.87 \\ \hline
1.225 & 0.913\hspace{0.5em}$\pm$ 0.012 & 176.25\hspace{0.5em}$\pm$ 57.10 & 0.550 & 1.067\hspace{0.5em}$\pm$ 0.022 & 126.33\hspace{0.5em}$\pm$ \phantom{0}8.00 & 0.112 & 17.94\hspace{0.5em}$\pm$ 1.02 \\
1.250 & 0.901\hspace{0.5em}$\pm$ 0.014 & 205.99\hspace{0.5em}$\pm$ 66.77 & 0.500 & 1.072\hspace{0.5em}$\pm$ 0.026 & 152.59\hspace{0.5em}$\pm$ \phantom{0}9.62 & 0.125 & 20.24\hspace{0.5em}$\pm$ 1.19 \\
1.275 & 0.889\hspace{0.5em}$\pm$ 0.015 & 241.13\hspace{0.5em}$\pm$ 78.19 & 0.450 & 1.074\hspace{0.5em}$\pm$ 0.026 & 187.32\hspace{0.5em}$\pm$ 11.97 & 0.137 & 22.64\hspace{0.5em}$\pm$ 1.37 \\ \hline
1.300 & 0.878\hspace{0.5em}$\pm$ 0.016 & 282.44\hspace{0.5em}$\pm$ 91.52 & 0.400 & 1.073\hspace{0.5em}$\pm$ 0.026 & 235.46\hspace{0.5em}$\pm$ 15.84 & 0.150 & 25.05\hspace{0.5em}$\pm$ 1.56 \\
\hline \hline
  \multicolumn{3}{|c||}{\sffamily{\bfseries{tensile modulus}}}
& \multicolumn{3}{c||} {\sffamily{\bfseries{compressive modulus}}}
& \multicolumn{2}{c|}  {\sffamily{\bfseries{shear modulus}}} \\
  \multicolumn{3}{|c||}{$\textsf{E}_{\rm{ten}} = 594.15 \pm 194.30$\,kPa}
& \multicolumn{3}{c||} {$\textsf{E}_{\rm{com}} = 312.93 \pm 64.49$\,kPa}
& \multicolumn{2}{c|}  {$\textsf{G} = 153.55 \pm 8.11$\,kPa} \\
\hline \hline
  \multicolumn{3}{|c||}{\sffamily{\bfseries{tensile Poisson's ratio}}}
& \multicolumn{3}{c||} {\sffamily{\bfseries{compressive Poisson's ratio}}}
& \multicolumn{2}{c|}{} \\
  \multicolumn{3}{|c||}{$\nu_{\rm{ten}} = 0.393 \pm 0.032$}
& \multicolumn{3}{c||} {$\nu_{\rm{com}} = 0.115 \pm 0.022$}
& \multicolumn{2}{c|}{} \\
\hline \hline
  \multicolumn{3}{|c||}{\sffamily{\bfseries{energy return}}}
& \multicolumn{3}{c||} {\sffamily{\bfseries{energy return}}}
& \multicolumn{2}{c|}  {\sffamily{\bfseries{energy return}}} \\
  \multicolumn{3}{|c||}{$\eta_{\rm{ten}}  = 88.9 \pm 1.4 \%$}
& \multicolumn{3}{c||} {$\eta_{\rm{com}}  = 86.6 \pm 0.2\%$}
& \multicolumn{2}{c|}  {$\eta_{\rm{shr}}  = 70.6 \pm 1.0 \%$} \\
\hline
\end{tabular}    
\vspace*{0.5cm}
\caption{{\sffamily{\bfseries{
Stress, transverse stretch, and mechanical properties of the midfoot region
of the worn shoe.}}}
Values report mean $\pm$ standard deviation for $n=5$ samples in tension,
compression, and shear, including the elastic modulus, Poisson's ratio, and energy return.}
    \label{tab:raw_data_worn_mid}
    \centering
    \setlength{\tabcolsep}{3pt}
            \fontsize{10pt}{11pt}\selectfont
\begin{tabular}{|ccc||ccc||cc|}
\hline
  \multicolumn{3}{|c||}{\sffamily{\bfseries{uniaxial tension}}}
& \multicolumn{3}{c||} {\sffamily{\bfseries{uniaxial compression}}}
& \multicolumn{2}{c|}  {\sffamily{\bfseries{simple shear}}} \\
  \multicolumn{3}{|c||}{$n=5$}
& \multicolumn{3}{c||}{$n=5$}
& \multicolumn{2}{c|}{$n=5$} \\ \hline
$\lambda_1$ & $\lambda_2$ & $P_{11}$ & $\lambda_1$ & $\lambda_2$ & $|P_{11}|$ & $\gamma$ & $P_{12}$ \\
\,[-] & [-] & [kPa] & [-] & [-] & [kPa] & [-] & [kPa] \\
\hline \hline
1.000 & 1.000\hspace{0.5em}$\pm$ 0.000 & \phantom{0}\phantom{0}0.00\hspace{0.5em}$\pm$ \phantom{0}0.00 & 1.000 & 1.000\hspace{0.5em}$\pm$ 0.000 & \phantom{0}\phantom{0}0.00\hspace{0.5em}$\pm$ \phantom{0}0.00 & 0.000 & \phantom{0}0.00\hspace{0.5em}$\pm$ 0.00 \\ \hline
1.025 & 0.995\hspace{0.5em}$\pm$ 0.003 & \phantom{0}11.09\hspace{0.5em}$\pm$ \phantom{0}0.90 & 0.950 & 1.008\hspace{0.5em}$\pm$ 0.001 & \phantom{0}\phantom{0}9.99\hspace{0.5em}$\pm$ \phantom{0}2.62 & 0.012 & \phantom{0}1.82\hspace{0.5em}$\pm$ 0.24 \\
1.050 & 0.989\hspace{0.5em}$\pm$ 0.006 & \phantom{0}26.12\hspace{0.5em}$\pm$ \phantom{0}2.11 & 0.900 & 1.014\hspace{0.5em}$\pm$ 0.001 & \phantom{0}28.16\hspace{0.5em}$\pm$ \phantom{0}5.06 & 0.025 & \phantom{0}3.76\hspace{0.5em}$\pm$ 0.50 \\
1.075 & 0.984\hspace{0.5em}$\pm$ 0.009 & \phantom{0}43.00\hspace{0.5em}$\pm$ \phantom{0}3.68 & 0.850 & 1.019\hspace{0.5em}$\pm$ 0.002 & \phantom{0}42.36\hspace{0.5em}$\pm$ \phantom{0}5.56 & 0.037 & \phantom{0}5.71\hspace{0.5em}$\pm$ 0.77 \\ \hline
1.100 & 0.976\hspace{0.5em}$\pm$ 0.011 & \phantom{0}62.47\hspace{0.5em}$\pm$ \phantom{0}5.69 & 0.800 & 1.023\hspace{0.5em}$\pm$ 0.003 & \phantom{0}52.92\hspace{0.5em}$\pm$ \phantom{0}6.17 & 0.050 & \phantom{0}7.68\hspace{0.5em}$\pm$ 1.05 \\ \hline
1.125 & 0.968\hspace{0.5em}$\pm$ 0.012 & \phantom{0}84.53\hspace{0.5em}$\pm$ \phantom{0}8.00 & 0.750 & 1.027\hspace{0.5em}$\pm$ 0.004 & \phantom{0}63.28\hspace{0.5em}$\pm$ \phantom{0}6.71 & 0.062 & \phantom{0}9.69\hspace{0.5em}$\pm$ 1.34 \\
1.150 & 0.960\hspace{0.5em}$\pm$ 0.014 & 108.78\hspace{0.5em}$\pm$ 10.37 & 0.700 & 1.030\hspace{0.5em}$\pm$ 0.005 & \phantom{0}74.98\hspace{0.5em}$\pm$ \phantom{0}7.28 & 0.075 & 11.73\hspace{0.5em}$\pm$ 1.63 \\
1.175 & 0.951\hspace{0.5em}$\pm$ 0.016 & 134.72\hspace{0.5em}$\pm$ 12.73 & 0.650 & 1.034\hspace{0.5em}$\pm$ 0.006 & \phantom{0}89.01\hspace{0.5em}$\pm$ \phantom{0}7.99 & 0.087 & 13.79\hspace{0.5em}$\pm$ 1.93 \\ \hline
1.200 & 0.941\hspace{0.5em}$\pm$ 0.017 & 162.37\hspace{0.5em}$\pm$ 15.13 & 0.600 & 1.038\hspace{0.5em}$\pm$ 0.007 & 106.14\hspace{0.5em}$\pm$ \phantom{0}8.98 & 0.100 & 15.99\hspace{0.5em}$\pm$ 2.29 \\ \hline
1.225 & 0.931\hspace{0.5em}$\pm$ 0.018 & 192.31\hspace{0.5em}$\pm$ 17.67 & 0.550 & 1.041\hspace{0.5em}$\pm$ 0.008 & 127.13\hspace{0.5em}$\pm$ 10.27 & 0.112 & 18.16\hspace{0.5em}$\pm$ 2.63 \\
1.250 & 0.921\hspace{0.5em}$\pm$ 0.020 & 225.62\hspace{0.5em}$\pm$ 20.55 & 0.500 & 1.046\hspace{0.5em}$\pm$ 0.008 & 154.01\hspace{0.5em}$\pm$ 12.11 & 0.125 & 20.44\hspace{0.5em}$\pm$ 2.99 \\
1.275 & 0.911\hspace{0.5em}$\pm$ 0.021 & 263.87\hspace{0.5em}$\pm$ 23.87 & 0.450 & 1.048\hspace{0.5em}$\pm$ 0.010 & 189.45\hspace{0.5em}$\pm$ 14.66 & 0.137 & 22.80\hspace{0.5em}$\pm$ 3.35 \\ \hline
1.300 & 0.902\hspace{0.5em}$\pm$ 0.022 & 307.48\hspace{0.5em}$\pm$ 27.92 & 0.400 & 1.049\hspace{0.5em}$\pm$ 0.011 & 238.73\hspace{0.5em}$\pm$ 19.34 & 0.150 & 25.14\hspace{0.5em}$\pm$ 3.73 \\
\hline \hline
  \multicolumn{3}{|c||}{\sffamily{\bfseries{tensile modulus}}}
& \multicolumn{3}{c||} {\sffamily{\bfseries{compressive modulus}}}
& \multicolumn{2}{c|}  {\sffamily{\bfseries{shear modulus}}} \\
  \multicolumn{3}{|c||}{$\textsf{E}_{\rm{ten}} = 572.66 \pm 49.27$\,kPa}
& \multicolumn{3}{c||} {$\textsf{E}_{\rm{com}} = 246.84 \pm 52.81$\,kPa}
& \multicolumn{2}{c|}  {$\textsf{G} = 156.47 \pm 21.86$\,kPa} \\
\hline \hline
  \multicolumn{3}{|c||}{\sffamily{\bfseries{tensile Poisson's ratio}}}
& \multicolumn{3}{c||} {\sffamily{\bfseries{compressive Poisson's ratio}}}
& \multicolumn{2}{c|}{} \\
  \multicolumn{3}{|c||}{$\nu_{\rm{ten}} = 0.333 \pm 0.044$}
& \multicolumn{3}{c||} {$\nu_{\rm{com}} = 0.179 \pm 0.067$}
& \multicolumn{2}{c|}{} \\
\hline \hline
  \multicolumn{3}{|c||}{\sffamily{\bfseries{energy return}}}
& \multicolumn{3}{c||} {\sffamily{\bfseries{energy return}}}
& \multicolumn{2}{c|}  {\sffamily{\bfseries{energy return}}} \\
  \multicolumn{3}{|c||}{$\eta_{\rm{ten}}  = 88.8 \pm 0.6 \%$}
& \multicolumn{3}{c||} {$\eta_{\rm{com}}  = 87.4 \pm 0.8\%$}
& \multicolumn{2}{c|}  {$\eta_{\rm{shr}}  = 68.0 \pm 1.5 \%$} \\
\hline
\end{tabular}    
\end{table*}

\begin{table*}[p]
\caption{{\sffamily{\bfseries{
Stress, transverse stretch, and mechanical properties of the heel region
of the new shoe.}}}
Values report mean $\pm$ standard deviation for $n=5$ samples in tension,
compression, and shear, including the elastic modulus, Poisson's ratio, and energy return.}
\label{tab:raw_data_new_heel}
    \centering
    \newcommand{\tablesize}{\footnotesize}
    \setlength{\tabcolsep}{3pt}
            \fontsize{10pt}{11pt}\selectfont
\begin{tabular}{|ccc||ccc||cc|}
\hline
  \multicolumn{3}{|c||}{\sffamily{\bfseries{uniaxial tension}}}
& \multicolumn{3}{c||} {\sffamily{\bfseries{uniaxial compression}}}
& \multicolumn{2}{c|}  {\sffamily{\bfseries{simple shear}}} \\
  \multicolumn{3}{|c||}{$n=5$}
& \multicolumn{3}{c||}{$n=5$}
& \multicolumn{2}{c|}{$n=5$} \\ \hline
$\lambda_1$ & $\lambda_2$ & $P_{11}$ & $\lambda_1$ & $\lambda_2$ & $|P_{11}|$ & $\gamma$ & $P_{12}$ \\
\,[-] & [-] & [kPa] & [-] & [-] & [kPa] & [-] & [kPa] \\
\hline \hline
1.000 & 1.000\hspace{0.5em}$\pm$ 0.000 & \phantom{0}\phantom{0}0.00\hspace{0.5em}$\pm$ \phantom{0}0.00 & 1.000 & 1.000\hspace{0.5em}$\pm$ 0.000 & \phantom{0}\phantom{0}0.00\hspace{0.5em}$\pm$ \phantom{0}0.00 & 0.000 & \phantom{0}0.00\hspace{0.5em}$\pm$ 0.00 \\ \hline
1.025 & 0.993\hspace{0.5em}$\pm$ 0.001 & \phantom{0}18.84\hspace{0.5em}$\pm$ \phantom{0}1.54 & 0.950 & 1.007\hspace{0.5em}$\pm$ 0.003 & \phantom{0}15.28\hspace{0.5em}$\pm$ \phantom{0}3.56 & 0.012 & \phantom{0}2.06\hspace{0.5em}$\pm$ 0.27 \\
1.050 & 0.985\hspace{0.5em}$\pm$ 0.002 & \phantom{0}41.72\hspace{0.5em}$\pm$ \phantom{0}2.87 & 0.900 & 1.014\hspace{0.5em}$\pm$ 0.003 & \phantom{0}38.69\hspace{0.5em}$\pm$ \phantom{0}5.04 & 0.025 & \phantom{0}4.25\hspace{0.5em}$\pm$ 0.55 \\
1.075 & 0.978\hspace{0.5em}$\pm$ 0.002 & \phantom{0}66.35\hspace{0.5em}$\pm$ \phantom{0}4.18 & 0.850 & 1.019\hspace{0.5em}$\pm$ 0.004 & \phantom{0}54.76\hspace{0.5em}$\pm$ \phantom{0}5.94 & 0.037 & \phantom{0}6.44\hspace{0.5em}$\pm$ 0.84 \\ \hline
1.100 & 0.968\hspace{0.5em}$\pm$ 0.002 & \phantom{0}93.11\hspace{0.5em}$\pm$ \phantom{0}5.48 & 0.800 & 1.023\hspace{0.5em}$\pm$ 0.004 & \phantom{0}66.30\hspace{0.5em}$\pm$ \phantom{0}6.54 & 0.050 & \phantom{0}8.65\hspace{0.5em}$\pm$ 1.12 \\ \hline
1.125 & 0.957\hspace{0.5em}$\pm$ 0.003 & 122.22\hspace{0.5em}$\pm$ \phantom{0}6.79 & 0.750 & 1.028\hspace{0.5em}$\pm$ 0.005 & \phantom{0}77.67\hspace{0.5em}$\pm$ \phantom{0}7.11 & 0.062 & 10.88\hspace{0.5em}$\pm$ 1.39 \\
1.150 & 0.946\hspace{0.5em}$\pm$ 0.004 & 153.06\hspace{0.5em}$\pm$ \phantom{0}8.22 & 0.700 & 1.032\hspace{0.5em}$\pm$ 0.005 & \phantom{0}90.79\hspace{0.5em}$\pm$ \phantom{0}7.65 & 0.075 & 13.14\hspace{0.5em}$\pm$ 1.65 \\
1.175 & 0.934\hspace{0.5em}$\pm$ 0.005 & 185.69\hspace{0.5em}$\pm$ \phantom{0}9.70 & 0.650 & 1.037\hspace{0.5em}$\pm$ 0.006 & 106.46\hspace{0.5em}$\pm$ \phantom{0}8.25 & 0.087 & 15.45\hspace{0.5em}$\pm$ 1.94 \\ \hline
1.200 & 0.922\hspace{0.5em}$\pm$ 0.007 & 220.96\hspace{0.5em}$\pm$ 11.23 & 0.600 & 1.041\hspace{0.5em}$\pm$ 0.007 & 125.70\hspace{0.5em}$\pm$ \phantom{0}9.06 & 0.100 & 17.76\hspace{0.5em}$\pm$ 2.15 \\ \hline
1.225 & 0.910\hspace{0.5em}$\pm$ 0.008 & 260.37\hspace{0.5em}$\pm$ 12.81 & 0.550 & 1.047\hspace{0.5em}$\pm$ 0.008 & 149.90\hspace{0.5em}$\pm$ 10.18 & 0.112 & 20.18\hspace{0.5em}$\pm$ 2.38 \\
1.250 & 0.898\hspace{0.5em}$\pm$ 0.008 & 305.96\hspace{0.5em}$\pm$ 14.48 & 0.500 & 1.052\hspace{0.5em}$\pm$ 0.009 & 180.91\hspace{0.5em}$\pm$ 11.68 & 0.125 & 22.57\hspace{0.5em}$\pm$ 2.64 \\
1.275 & 0.887\hspace{0.5em}$\pm$ 0.009 & 360.45\hspace{0.5em}$\pm$ 16.28 & 0.450 & 1.061\hspace{0.5em}$\pm$ 0.012 & 222.71\hspace{0.5em}$\pm$ 14.06 & 0.137 & 25.05\hspace{0.5em}$\pm$ 2.79 \\ \hline
1.300 & 0.876\hspace{0.5em}$\pm$ 0.010 & 425.34\hspace{0.5em}$\pm$ 18.48 & 0.400 & 1.062\hspace{0.5em}$\pm$ 0.012 & 281.94\hspace{0.5em}$\pm$ 17.64 & 0.150 & 27.47\hspace{0.5em}$\pm$ 2.95 \\
\hline \hline
  \multicolumn{3}{|c||}{\sffamily{\bfseries{tensile modulus}}}
& \multicolumn{3}{c||} {\sffamily{\bfseries{compressive modulus}}}
& \multicolumn{2}{c|}  {\sffamily{\bfseries{shear modulus}}} \\
  \multicolumn{3}{|c||}{$\textsf{E}_{\rm{ten}} = 882.49 \pm 56.01$\,kPa}
& \multicolumn{3}{c||} {$\textsf{E}_{\rm{com}} = 353.91 \pm 59.09$\,kPa}
& \multicolumn{2}{c|}  {$\textsf{G} = 175.21 \pm 22.05$\,kPa} \\
\hline \hline
  \multicolumn{3}{|c||}{\sffamily{\bfseries{tensile Poisson's ratio}}}
& \multicolumn{3}{c||} {\sffamily{\bfseries{compressive Poisson's ratio}}}
& \multicolumn{2}{c|}{} \\
  \multicolumn{3}{|c||}{$\nu_{\rm{ten}} = 0.302 \pm 0.083$}
& \multicolumn{3}{c||} {$\nu_{\rm{com}} = 0.112 \pm 0.014$}
& \multicolumn{2}{c|}{} \\
\hline \hline
  \multicolumn{3}{|c||}{\sffamily{\bfseries{energy return}}}
& \multicolumn{3}{c||} {\sffamily{\bfseries{energy return}}}
& \multicolumn{2}{c|}  {\sffamily{\bfseries{energy return}}} \\
  \multicolumn{3}{|c||}{$\eta_{\rm{ten}}  = 92.8 \pm 1.4 \%$}
& \multicolumn{3}{c||} {$\eta_{\rm{com}}  = 88.1 \pm 1.0\%$}
& \multicolumn{2}{c|}  {$\eta_{\rm{shr}}  = 63.6 \pm 7.6 \%$} \\
\hline
\end{tabular}
\vspace*{0.5cm}
\caption{{\sffamily{\bfseries{
Stress, transverse stretch, and mechanical properties of the heel region
of the worn shoe.}}}
Values report mean $\pm$ standard deviation for $n=5$ samples in tension,
compression, and shear, including the elastic modulus, Poisson's ratio, and energy return.}
    \label{tab:raw_data_worn_heel} 
    \centering 
    \setlength{\tabcolsep}{3pt}
            \fontsize{10pt}{11pt}\selectfont 
\begin{tabular}{|ccc||ccc||cc|}
\hline
  \multicolumn{3}{|c||}{\sffamily{\bfseries{uniaxial tension}}}
& \multicolumn{3}{c||} {\sffamily{\bfseries{uniaxial compression}}}
& \multicolumn{2}{c|}  {\sffamily{\bfseries{simple shear}}} \\
  \multicolumn{3}{|c||}{$n=5$}
& \multicolumn{3}{c||}{$n=5$}
& \multicolumn{2}{c|}{$n=5$} \\ \hline
$\lambda_1$ & $\lambda_2$ & $P_{11}$ & $\lambda_1$ & $\lambda_2$ & $|P_{11}|$ & $\gamma$ & $P_{12}$ \\
\,[-] & [-] & [kPa] & [-] & [-] & [kPa] & [-] & [kPa] \\
\hline \hline
1.000 & 1.000\hspace{0.5em}$\pm$ 0.000 & \phantom{0}\phantom{0}0.00\hspace{0.5em}$\pm$ \phantom{0}0.00 & 1.000 & 1.000\hspace{0.5em}$\pm$ 0.000 & \phantom{0}\phantom{0}0.00\hspace{0.5em}$\pm$ \phantom{0}0.00 & 0.000 & \phantom{0}0.00\hspace{0.5em}$\pm$ 0.00 \\ \hline
1.025 & 0.997\hspace{0.5em}$\pm$ 0.001 & \phantom{0}14.39\hspace{0.5em}$\pm$ \phantom{0}1.59 & 0.950 & 1.008\hspace{0.5em}$\pm$ 0.003 & \phantom{0}12.52\hspace{0.5em}$\pm$ \phantom{0}2.18 & 0.012 & \phantom{0}2.10\hspace{0.5em}$\pm$ 0.36 \\
1.050 & 0.992\hspace{0.5em}$\pm$ 0.000 & \phantom{0}33.84\hspace{0.5em}$\pm$ \phantom{0}3.77 & 0.900 & 1.018\hspace{0.5em}$\pm$ 0.003 & \phantom{0}33.04\hspace{0.5em}$\pm$ \phantom{0}5.11 & 0.025 & \phantom{0}4.33\hspace{0.5em}$\pm$ 0.73 \\
1.075 & 0.987\hspace{0.5em}$\pm$ 0.001 & \phantom{0}55.99\hspace{0.5em}$\pm$ \phantom{0}6.50 & 0.850 & 1.026\hspace{0.5em}$\pm$ 0.004 & \phantom{0}47.96\hspace{0.5em}$\pm$ \phantom{0}7.52 & 0.037 & \phantom{0}6.56\hspace{0.5em}$\pm$ 1.11 \\ \hline
1.100 & 0.980\hspace{0.5em}$\pm$ 0.002 & \phantom{0}81.14\hspace{0.5em}$\pm$ \phantom{0}9.56 & 0.800 & 1.032\hspace{0.5em}$\pm$ 0.004 & \phantom{0}59.47\hspace{0.5em}$\pm$ \phantom{0}8.85 & 0.050 & \phantom{0}8.81\hspace{0.5em}$\pm$ 1.48 \\ \hline
1.125 & 0.972\hspace{0.5em}$\pm$ 0.002 & 108.87\hspace{0.5em}$\pm$ 12.67 & 0.750 & 1.035\hspace{0.5em}$\pm$ 0.005 & \phantom{0}70.67\hspace{0.5em}$\pm$ \phantom{0}9.79 & 0.062 & 11.08\hspace{0.5em}$\pm$ 1.87 \\
1.150 & 0.962\hspace{0.5em}$\pm$ 0.003 & 138.52\hspace{0.5em}$\pm$ 15.64 & 0.700 & 1.040\hspace{0.5em}$\pm$ 0.006 & \phantom{0}83.20\hspace{0.5em}$\pm$ 10.77 & 0.075 & 13.39\hspace{0.5em}$\pm$ 2.27 \\
1.175 & 0.952\hspace{0.5em}$\pm$ 0.003 & 169.84\hspace{0.5em}$\pm$ 18.65 & 0.650 & 1.044\hspace{0.5em}$\pm$ 0.007 & \phantom{0}98.05\hspace{0.5em}$\pm$ 11.79 & 0.087 & 15.69\hspace{0.5em}$\pm$ 2.62 \\ \hline
1.200 & 0.940\hspace{0.5em}$\pm$ 0.003 & 203.32\hspace{0.5em}$\pm$ 21.81 & 0.600 & 1.046\hspace{0.5em}$\pm$ 0.009 & 116.16\hspace{0.5em}$\pm$ 13.13 & 0.100 & 18.11\hspace{0.5em}$\pm$ 3.02 \\ \hline
1.225 & 0.929\hspace{0.5em}$\pm$ 0.004 & 239.98\hspace{0.5em}$\pm$ 25.40 & 0.550 & 1.048\hspace{0.5em}$\pm$ 0.012 & 138.46\hspace{0.5em}$\pm$ 15.03 & 0.112 & 20.46\hspace{0.5em}$\pm$ 3.36 \\
1.250 & 0.917\hspace{0.5em}$\pm$ 0.004 & 281.52\hspace{0.5em}$\pm$ 29.61 & 0.500 & 1.051\hspace{0.5em}$\pm$ 0.016 & 167.02\hspace{0.5em}$\pm$ 17.50 & 0.125 & 22.91\hspace{0.5em}$\pm$ 3.71 \\
1.275 & 0.906\hspace{0.5em}$\pm$ 0.005 & 330.06\hspace{0.5em}$\pm$ 34.87 & 0.450 & 1.053\hspace{0.5em}$\pm$ 0.016 & 204.49\hspace{0.5em}$\pm$ 21.25 & 0.137 & 25.40\hspace{0.5em}$\pm$ 4.13 \\ \hline
1.300 & 0.895\hspace{0.5em}$\pm$ 0.005 & 386.76\hspace{0.5em}$\pm$ 41.49 & 0.400 & 1.054\hspace{0.5em}$\pm$ 0.016 & 257.47\hspace{0.5em}$\pm$ 28.09 & 0.150 & 27.82\hspace{0.5em}$\pm$ 4.57 \\
\hline \hline
  \multicolumn{3}{|c||}{\sffamily{\bfseries{tensile modulus}}}
& \multicolumn{3}{c||} {\sffamily{\bfseries{compressive modulus}}}
& \multicolumn{2}{c|}  {\sffamily{\bfseries{shear modulus}}} \\
  \multicolumn{3}{|c||}{$\textsf{E}_{\rm{ten}} = 744.54 \pm 86.11$\,kPa}
& \multicolumn{3}{c||} {$\textsf{E}_{\rm{com}} = 298.02 \pm 45.83$\,kPa}
& \multicolumn{2}{c|}  {$\textsf{G} = 178.48 \pm 30.04$\,kPa} \\
\hline \hline
  \multicolumn{3}{|c||}{\sffamily{\bfseries{tensile Poisson's ratio}}}
& \multicolumn{3}{c||} {\sffamily{\bfseries{compressive Poisson's ratio}}}
& \multicolumn{2}{c|}{} \\
  \multicolumn{3}{|c||}{$\nu_{\rm{ten}} = 0.309 \pm 0.017$}
& \multicolumn{3}{c||} {$\nu_{\rm{com}} = 0.149 \pm 0.021$}
& \multicolumn{2}{c|}{} \\
\hline \hline
  \multicolumn{3}{|c||}{\sffamily{\bfseries{energy return}}}
& \multicolumn{3}{c||} {\sffamily{\bfseries{energy return}}}
& \multicolumn{2}{c|}  {\sffamily{\bfseries{energy return}}} \\
  \multicolumn{3}{|c||}{$\eta_{\rm{ten}}  = 91.2 \pm 0.5 \%$}
& \multicolumn{3}{c||} {$\eta_{\rm{com}}  = 88.5 \pm 0.5\%$}
& \multicolumn{2}{c|}  {$\eta_{\rm{shr}}  = 66.4 \pm 1.9 \%$} \\
\hline
\end{tabular}
\end{table*}
\bibliographystyle{cas-model2-names}
\bibliography{references}

@article{verdejo_2004,
  author  = {Verdejo, R. and Mills, N. J.},
  title   = {Simulating the effects of long distance running on shoe midsole foam},
  journal = {Polymer Testing},
  volume  = {23},
  number  = {5},
  pages   = {567--574},
  year    = {2004},
  doi     = {10.1016/j.polymertesting.2003.11.005}
}

@article{aimar_2023,
  author  = {Aimar, C. and Org{\'e}as, L. and Rolland du Roscoat, S. and Bailly, L. and Ferr{\'e} Sentis, D.},
  title   = {Fatigue mechanisms of a closed cell elastomeric foam: A mechanical and microstructural study using ex situ {X}-ray microtomography},
  journal = {Polymer Testing},
  volume  = {128},
  pages   = {108194},
  year    = {2023},
  doi     = {10.1016/j.polymertesting.2023.108194}
}

@article{fakayode_2025,
  author  = {Fakayode, Sayo O. and Rosado Flores, Peter and Bolton, Brinkley and Dassow, Bailey and Moore, Kate and Owens, Kayley},
  title   = {Testing and analysis of footwear insole copolymers by {FTIR-ATR}, non-sampling {Raman} probe spectroscopy, and thermal gravimetric analysis},
  journal = {Polymer Testing},
  volume  = {149},
  pages   = {108847},
  year    = {2025},
  doi     = {10.1016/j.polymertesting.2025.108847}
}

@article{lunchev_2022,
  author  = {Lunchev, Andrey V. and Kashcheev, Aleksandr and Tok, Alfred Ling Yoong and Lipik, Vitali},
  title   = {Mechanical characteristics of poly(ethylene vinyl acetate) foams with graphene for the applications in sport footwear},
  journal = {Polymer Testing},
  volume  = {113},
  pages   = {107688},
  year    = {2022},
  doi     = {10.1016/j.polymertesting.2022.107688}
}

@article{rodrigo-carranza_influence_2024,
	title = {Influence of different midsole foam in advanced footwear technology use on running economy and biomechanics in trained runners},
	volume = {34},
	url = {https://onlinelibrary.wiley.com/doi/abs/10.1111/sms.14526},
	doi = {https://doi.org/10.1111/sms.14526},
	number = {1},
	journal = {Scandinavian Journal of Medicine \& Science in Sports},
	author = {Rodrigo-Carranza, Víctor and Hoogkamer, Wouter and González-Ravé, José María and Horta-Muñoz, Sergio and Serna-Moreno, María del Carmen and Romero-Gutierrez, Ana and González-Mohíno, Fernando},
	year = {2024},
	pages = {e14526},
}

@article{rosenberg_what_2022,
	title = {What money can buy: technology and breaking the two-hour 'marathon' record},
	volume = {49},
	issn = {0094-8705},
	url = {https://doi.org/10.1080/00948705.2021.1976194},
	doi = {10.1080/00948705.2021.1976194},
	number = {1},
	journal = {Journal of the Philosophy of Sport},
	publisher = {Routledge},
	author = {Rosenberg, Danny and Sailors, Pam R.},
	month = jan,
	year = {2022},
	pages = {1--18},
}

@article{kirby_influence_2019,
	title = {Influence of performance running footwear on muscle soreness and damage},
	volume = {11},
	url = {https://doi.org/10.1080/19424280.2019.1606325},
	doi = {10.1080/19424280.2019.1606325},
	number = {sup1},
	journal = {Footwear Science},
	publisher = {Taylor \& Francis},
	author = {Kirby, Brett Sean and Hughes, Elizabeth and Haines, Michelle and Stinman, Sarah and Winn, Brad J.},
	year = {2019},
	pages = {S188--S189},
}

@article{welch_generalization_1947,
	title = {The generalization of ''{Student}'s'' problem when several different population variances are involved},
	volume = {34},
	issn = {0006-3444},
	url = {https://doi.org/10.1093/biomet/34.1-2.28},
	doi = {10.1093/biomet/34.1-2.28},
	number = {1-2},
	urldate = {2026-08-24},
	journal = {Biometrika},
	author = {Welch, B. L.},
	month = jan,
	year = {1947},
	pages = {28--35},
}

@article{bruvere_mechanisms_2025,
	title = {Mechanisms, {Economy}, and {Performance} of {Advanced} {Footwear} {Technology} in {Endurance} {Running}---{A} {Review}},
	volume = {5},
	number = {1},
	journal = {Muscles},
	author = {Bruvere, Daido Dagne and Bernans, Edgars},
	year = {2025},
	pages = {2},
}

@misc{deep_market_insights_carbon_2026,
	title = {Carbon {Plate} {Running} {Shoes} {Market} {Size}, {Trends} \& {Global} {Demand} {\textbar} 2031},
	url = {https://deepmarketinsights.com/report/carbon-plate-running-shoes-market-research-report},
	language = {en},
	urldate = {2026-07-17},
	author = {Deep Market Insights},
	month = jun,
	year = {2026},
}

@article{wang_durability_2012,
	title = {Durability of running shoes with ethylene vinyl acetate or polyurethane midsoles},
	volume = {30},
	issn = {0264-0414},
	url = {https://doi.org/10.1080/02640414.2012.723819},
	doi = {10.1080/02640414.2012.723819},
	number = {16},
	journal = {Journal of Sports Sciences},
	publisher = {Routledge},
	author = {Wang, Lin and Hong, Youlian and Li, Jing Xian},
	month = dec,
	year = {2012},
	pages = {1787--1792},
}

@article{wang_changes_2010,
	title = {Changes in heel cushioning characteristics of running shoes with running mileage},
	volume = {2},
	issn = {1942-4280},
	url = {https://doi.org/10.1080/19424280.2010.519348},
	doi = {10.1080/19424280.2010.519348},
	number = {3},
	journal = {Footwear Science},
	publisher = {Taylor \& Francis},
	author = {Wang, Lin and Xian Li, Jing and Hong, Youlian and He Zhou, Ji},
	month = sep,
	year = {2010},
	pages = {141--147},
}

@article{kuzmeski_data_2026,
	title = {Data driven shoe design improves running economy beyond state-of-the-art {Advanced} {Footwear} {Technology} running shoes},
	volume = {15},
	issn = {2095-2546},
	url = {https://www.sciencedirect.com/science/article/pii/S2095254626000116},
	doi = {10.1016/j.jshs.2026.101133},
	journal = {Journal of Sport and Health Science},
	author = {Kuzmeski, John and Bertschy, Montgomery and Healey, Laura and Barrons, Zach and Hoogkamer, Wouter},
	month = dec,
	year = {2026},
	pages = {101133},
}

@article{chapman_ground_2012,
	title = {Ground {Contact} {Time} as an {Indicator} of {Metabolic} {Cost} in {Elite} {Distance} {Runners}},
	volume = {44},
	issn = {1530-0315},
	doi = {10.1249/MSS.0b013e3182400520},
	number = {5},
	journal = {Medicine \& Science in Sports \& Exercise},
	publisher = {Wolters Kluwer Health \_ Lippincott Williams \& Wilkins},
	author = {Chapman, Rober and Laymon, Abigail and Whilhite, Daniel and McKenzie, James and Tanner, David and Stager, Joel},
	month = may,
	year = {2012},
	pages = {917--925},
}

@article{holm_simple_1979,
	title = {A {Simple} {Sequentially} {Rejective} {Multiple} {Test} {Procedure}},
	volume = {6},
	issn = {0303-6898},
	url = {https://www.jstor.org/stable/4615733},
	number = {2},
	urldate = {2026-08-04},
	journal = {Scandinavian Journal of Statistics},
	publisher = {[Board of the Foundation of the Scandinavian Journal of Statistics, Wiley]},
	author = {Holm, Sture},
	year = {1979},
	pages = {65--70},
}

@misc{boston_athletic_association_qualify_nodate,
	title = {Qualify for the {Boston} {Marathon}},
	url = {https://www.baa.org/races/boston-marathon/qualify/},
	language = {en-US},
	urldate = {2026-07-16},
	journal = {Boston Athletic Association},
	author = {{Boston Athletic Association}},
}

@article{cheah_manufacturing-focused_2013,
	title = {Manufacturing-focused emissions reductions in footwear production},
	volume = {44},
	issn = {0959-6526},
	url = {https://www.sciencedirect.com/science/article/pii/S0959652612006300},
	doi = {10.1016/j.jclepro.2012.11.037},
	journal = {Journal of Cleaner Production},
	author = {Cheah, Lynette and Ciceri, Natalia Duque and Olivetti, Elsa and Matsumura, Seiko and Forterre, Dai and Roth, Richard and Kirchain, Randolph},
	month = apr,
	year = {2013},
	pages = {18--29},
}

@misc{ou_multi-block_2026,
	title = {Multi-block copolyester ether thermoplastic elastomer foam, preparation method thereof, and sports shoe midsole made from the same},
	publisher = {Google Patents},
	author = {Ou, Fu-Wen and Lin, Ming-Champ and Wu, Jia-Ying},
	month = jan,
	year = {2026},
}

@article{whiting_metabolic_2022,
	title = {Metabolic cost of level, uphill, and downhill running in highly cushioned shoes with carbon-fiber plates},
	volume = {11},
	issn = {2095-2546},
	url = {https://www.sciencedirect.com/science/article/pii/S2095254621001113},
	doi = {10.1016/j.jshs.2021.10.004},
	number = {3},
	journal = {Journal of Sport and Health Science},
	author = {Whiting, Clarissa S. and Hoogkamer, Wouter and Kram, Rodger},
	month = may,
	year = {2022},
	pages = {303--308},
}

@article{mabe_changepoint_2024,
	title = {Changepoint {Analysis} on {Marathon} and {Half} {Marathon} {Data}: {An} {Insight} on {Annual} {Record} {Trends} and {Recent} {Major} {Marathon} {Performances}},
	author = {Mabe, Abigail},
	year = {2024},
}

@misc{mcculloch_discovering_2026,
	title = {Discovering the mechanics of ultra-low density elastomeric foams in elite-level racing shoes},
	url = {http://arxiv.org/abs/2602.12694},
	doi = {10.48550/arXiv.2602.12694},
	urldate = {2026-06-09},
	publisher = {arXiv},
	author = {McCulloch, Jeremy A. and Delp, Scott L. and Kuhl, Ellen},
	month = feb,
	year = {2026},
	note = {arXiv:2602.12694 [cs.CE]},
}

@article{grivas_why_2026,
	title = {Why now? {A} physiological perspective on the first official sub-2-hour marathon},
	issn = {1439-6327},
	url = {https://doi.org/10.1007/s00421-026-06285-8},
	doi = {10.1007/s00421-026-06285-8},
	journal = {European Journal of Applied Physiology},
	author = {Grivas, Gerasimos V.},
	month = jun,
	year = {2026},
}

@article{hoogkamer_comparison_2018,
	title = {A {Comparison} of the {Energetic} {Cost} of {Running} in {Marathon} {Racing} {Shoes}},
	volume = {48},
	issn = {1179-2035},
	doi = {10.1007/s40279-017-0811-2},
	language = {eng},
	number = {4},
	journal = {Sports Medicine (Auckland, N.Z.)},
	author = {Hoogkamer, Wouter and Kipp, Shalaya and Frank, Jesse H. and Farina, Emily M. and Luo, Geng and Kram, Rodger},
	month = apr,
	year = {2018},
	pages = {1009--1019},
}
\end{document}